\documentclass[12pt]{article}
\usepackage{amssymb}

\newtheorem{theorem}{Theorem}[section]
\newtheorem{corollary}{Corollary}[section]
\newtheorem{lemma}{Lemma}[section]

\begin{document}

\author{C. Bizdadea\thanks{%
e-mail address: bizdadea@central.ucv.ro}, C. C. Ciob\^\i rc\u {a}\thanks{%
e-mail address: ciobarca@central.ucv.ro}, E. M. Cioroianu\thanks{%
e-mail address: manache@central.ucv.ro}, \and S. O. Saliu\thanks{%
e-mail address: osaliu@central.ucv.ro}, S. C. S\u {a}raru\thanks{%
e-mail address: scsararu@central.ucv.ro} \\
Faculty of Physics, University of Craiova\\
13 A. I. Cuza Str., Craiova 200585, Romania}
\title{BRST cohomological results on the massless tensor field with the mixed
symmetry of the Riemann tensor}
\maketitle

\begin{abstract}
The basic BRST cohomological properties of a free, massless tensor field
with the mixed symmetry of the Riemann tensor are studied in detail. It is
shown that any non-trivial co-cycle from the local BRST cohomology group can
be taken to stop at antighost number three, its last component belonging to
the cohomology of the exterior longitudinal derivative and containing
non-trivial elements from the (invariant) characteristic cohomology.

\end{abstract}

\section{Introduction}

Tensor fields in ``exotic'' representations of the Lorentz group,
characterized by a mixed Young symmetry type~\cite
{curt,aul,labast,burd,zinov1}, appear in the context of many physically
interesting theories, like superstrings, supergravities or supersymmetric
high spin theories. Such models held the attention lately on some important
issues, like the dual formulation of field theories of spin two or higher~%
\cite{dualsp1,dualsp2,dualsp2a,dualsp3,dualsp4,dualsp5}, the impossibility
of consistent cross-interactions in the dual formulation of linearized
gravity~\cite{lingr} or a Lagrangian first-order approach~\cite
{zinov2,zinov3} to some classes of massless or partially massive mixed
symmetry-type tensor gauge fields, suggestively resembling to the tetrad
formalism of General Relativity. An important matter related to mixed
symmetry-type tensor fields is the study of their local BRST cohomology,
since it is helpful at determining the consistent interactions~\cite{def},
among themselves, as well as with higher-spin gauge theories~\cite
{high1,high2,high3,high4,noijhep}. The purpose of this paper is to
investigate the basic cohomological ingredients involved in the structure of
the co-cycles from the local BRST cohomology in the case of a free, massless
tensor gauge field $t_{\mu \nu |\alpha \beta }$ with the mixed symmetry of
the Riemann tensor .

More precisely, we initially determine the associated free antifield-BRST
symmetry $s$, which splits as the sum between the Koszul-Tate differential
and the exterior longitudinal derivative only, $s=\delta +\gamma $. Next, we
explicitly compute the cohomology of the exterior longitudinal derivative $%
H\left( \gamma \right) $ for the model under study and analyze other related
matters, like the triviality of the cohomology of the exterior spacetime
differential $d$ in the space of invariant polynomials and in $H\left(
\gamma \right) $. Further, we analyze the basic properties of the
characteristic cohomology (local cohomology of the Koszul-Tate differential)
$H\left( \delta |d\right) $ in pure ghost number zero and strictly positive
antighost number, as well as those of the invariant characteristic
cohomology $H^{\mathrm{inv}}\left( \delta |d\right) $. Finally, we consider
an arbitrary co-cycle from the local BRST cohomology $H\left( s|d\right) $,
of definite ghost number and in maximal form degree and show that if we
develop it along the antighost number, then one can always remove its
components of antighost number strictly greater than three by trivial
redefinitions only, while its non-trivial piece of highest antighost number
can always be taken to belong to $H\left( \gamma \right) $, with some
coefficients that are non-trivial elements from $H^{\mathrm{inv}}\left(
\delta |d\right) $. The results contained in this paper will be used at the
determination of consistent interactions for a free, massless tensor field
with the mixed symmetry of the Riemann tensor, alone and with other gauge
fields.

\section{Free model\label{2}: Lagrangian, gauge transformations and BRST
symmetry}

The starting point is given by the free Lagrangian action
\begin{eqnarray}
S_{0}\left[ t_{\mu \nu |\alpha \beta }\right] &=&\int d^{D}x\left( \frac{1}{8%
}\left( \partial ^{\lambda }t^{\mu \nu |\alpha \beta }\right) \left(
\partial _{\lambda }t_{\mu \nu |\alpha \beta }\right) -\frac{1}{2}\left(
\partial _{\mu }t^{\mu \nu |\alpha \beta }\right) \left( \partial ^{\lambda
}t_{\lambda \nu |\alpha \beta }\right) \right.  \nonumber \\
&&-\left( \partial _{\mu }t^{\mu \nu |\alpha \beta }\right) \left( \partial
_{\beta }t_{\nu \alpha }\right) -\frac{1}{2}\left( \partial ^{\lambda
}t^{\nu \beta }\right) \left( \partial _{\lambda }t_{\nu \beta }\right)
\nonumber \\
&&\left. +\left( \partial _{\nu }t^{\nu \beta }\right) \left( \partial
^{\lambda }t_{\lambda \beta }\right) -\frac{1}{2}\left( \partial _{\nu
}t^{\nu \beta }\right) \left( \partial _{\beta }t\right) +\frac{1}{8}\left(
\partial ^{\lambda }t\right) \left( \partial _{\lambda }t\right) \right) ,
\label{r1}
\end{eqnarray}
in a Minkowski-flat spacetime of dimension $D\geq 5$, endowed with a metric
tensor of `mostly plus' signature $\sigma _{\mu \nu }=\sigma ^{\mu \nu
}=\left( -++++\cdots \right) $. The massless tensor field $t_{\mu \nu
|\alpha \beta }$ of degree four has the mixed symmetry of the linearized
Riemann tensor, and hence transforms according to an irreducible
representation of $GL\left( D,\mathbb{R}\right) $, corresponding to the
rectangular Young diagram $\left( 2,2\right) $ with two columns and two
rows. Thus, it is separately antisymmetric in the pairs $\left\{ \mu ,\nu
\right\} $ and $\left\{ \alpha ,\beta \right\} $, is symmetric under the
interchange of these pairs ($\left\{ \mu ,\nu \right\} \longleftrightarrow
\left\{ \alpha ,\beta \right\} $) and satisfies the identity
\begin{equation}
t_{\left[ \mu \nu |\alpha \right] \beta }\equiv 0  \label{r5}
\end{equation}
associated with the above diagram, which we will refer to as the Bianchi I
identity. Here and in the sequel the symbol $\left[ \mu \nu \cdots \right] $
denotes the operation of antisymmetrization with respect to the indices
between brackets, without normalization factors. (For instance, the
left-hand side of (\ref{r5}) contains only three terms $t_{\left[ \mu \nu
|\alpha \right] \beta }=t_{\mu \nu |\alpha \beta }+t_{\nu \alpha |\mu \beta
}+t_{\alpha \mu |\nu \beta }$.) The notation $t_{\nu \beta }$ signifies the
simple trace of the original tensor field, which is symmetric, $t_{\nu \beta
}=\sigma ^{\mu \alpha }t_{\mu \nu |\alpha \beta }$, while $t$ denotes its
double trace, which is a scalar $t=\sigma ^{\nu \beta }t_{\nu \beta }$.

A generating set of gauge transformations for the action (\ref{r1}) reads as
\begin{equation}
\delta _{\epsilon }t_{\mu \nu |\alpha \beta }=\partial _{\mu }\epsilon
_{\alpha \beta |\nu }-\partial _{\nu }\epsilon _{\alpha \beta |\mu
}+\partial _{\alpha }\epsilon _{\mu \nu |\beta }-\partial _{\beta }\epsilon
_{\mu \nu |\alpha },  \label{r8}
\end{equation}
with the bosonic gauge parameters $\epsilon _{\mu \nu |\alpha }$
transforming according to an irreducible representation of $GL\left( D,%
\mathbb{R}\right) $, corresponding to the Young diagram $\left( 2,1\right) $
with two columns and two rows, being therefore antisymmetric in the pair $%
\mu \nu $ and satisfying the identity
\begin{equation}
\epsilon _{\left[ \mu \nu |\alpha \right] }\equiv 0.  \label{r10}
\end{equation}
The identity (\ref{r10}) is required in order to ensure that the gauge
transformations (\ref{r8}) check the same Bianchi I identity like the fields
themselves, namely, $\delta _{\epsilon }t_{\left[ \mu \nu |\alpha \right]
\beta }\equiv 0$. The above generating set of gauge transformations is
abelian and off-shell first-stage reducible since if we make the
transformation
\begin{equation}
\epsilon _{\mu \nu |\alpha }=2\partial _{\alpha }\theta _{\mu \nu }-\partial
_{\left[ \mu \right. }\theta _{\left. \nu \right] \alpha },  \label{r12}
\end{equation}
with $\theta _{\mu \nu }$ an arbitrary antisymmetric tensor ($\theta _{\mu
\nu }=-\theta _{\nu \mu }$), then the gauge transformations of the tensor
field identically vanish, $\delta _{\epsilon }t_{\mu \nu |\alpha \beta
}\equiv 0$. In the meantime, the transformation (\ref{r12}) is in agreement
with the identity (\ref{r10}) checked by the gauge parameters.

The field equations resulting from the action (\ref{r1}) take the form
\begin{equation}
\frac{\delta S_{0}}{\delta t^{\mu \nu |\alpha \beta }}\equiv -\frac{1}{4}%
T_{\mu \nu |\alpha \beta }\approx 0,  \label{r15}
\end{equation}
where $T_{\mu \nu |\alpha \beta }$ displays the same mixed symmetry
properties like the tensor field $t_{\mu \nu |\alpha \beta }$, is linear in
the field and second-order in its derivatives. Obviously, its simple trace $%
\sigma ^{\mu \alpha }T_{\mu \nu |\alpha \beta }\equiv T_{\nu \beta }$ is a
symmetric tensor, while its double trace is a scalar. The gauge invariance
of the Lagrangian action (\ref{r1}) under the transformations (\ref{r8}) is
equivalent to the fact that the functions defining the field equations are
not all independent, but rather obey the Noether identities
\begin{equation}
\partial ^{\mu }\frac{\delta S_{0}}{\delta t^{\mu \nu |\alpha \beta }}\equiv
-\frac{1}{4}\partial ^{\mu }T_{\mu \nu |\alpha \beta }=0,  \label{r17}
\end{equation}
while the first-stage reducibility shows that not all of the above Noether
identities are independent. It can be checked that the Euler-Lagrange (E.L.)
derivatives of the action, the gauge generators, as well as the first-order
reducibility functions, satisfy the general regularity assumptions from~\cite
{genreg}, such that the model under discussion is described by a normal
gauge theory of Cauchy order equal to three.

The most general gauge invariant quantities constructed out of the field $%
t_{\mu \nu |\alpha \beta }$ and its derivatives are given by the curvature
tensor
\begin{eqnarray}
F_{\mu \nu \lambda |\alpha \beta \gamma } &=&\partial _{\lambda }\partial
_{\gamma }t_{\mu \nu |\alpha \beta }+\partial _{\mu }\partial _{\gamma
}t_{\nu \lambda |\alpha \beta }+\partial _{\nu }\partial _{\gamma
}t_{\lambda \mu |\alpha \beta }  \nonumber \\
&&+\partial _{\lambda }\partial _{\alpha }t_{\mu \nu |\beta \gamma
}+\partial _{\mu }\partial _{\alpha }t_{\nu \lambda |\beta \gamma }+\partial
_{\nu }\partial _{\alpha }t_{\lambda \mu |\beta \gamma }  \nonumber \\
&&+\partial _{\lambda }\partial _{\beta }t_{\mu \nu |\gamma \alpha
}+\partial _{\mu }\partial _{\beta }t_{\nu \lambda |\gamma \alpha }+\partial
_{\nu }\partial _{\beta }t_{\lambda \mu |\gamma \alpha },  \label{curv}
\end{eqnarray}
together with its derivatives. As defined in the above, the curvature tensor
transforms in an irreducible representation of $GL\left( D,\mathbb{R}\right)
$ and exhibits the symmetries of the rectangular Young diagram $\left(
3,3\right) $ with two columns and three rows, so it is separately
antisymmetric in the indices $\left\{ \mu ,\nu ,\lambda \right\} $ and $%
\left\{ \alpha ,\beta ,\gamma \right\} $, symmetric under the interchange $%
\left\{ \mu ,\nu ,\lambda \right\} \longleftrightarrow \left\{ \alpha ,\beta
,\gamma \right\} $, and obeys the (algebraic) Bianchi I identity
\begin{equation}
F_{\left[ \mu \nu \lambda |\alpha \right] \beta \gamma }\equiv 0.
\label{r22}
\end{equation}
In addition, it verifies the (differential) Bianchi II identity
\begin{equation}
\partial _{\left[ \rho \right. }F_{\left. \mu \nu \lambda \right] |\alpha
\beta \gamma }\equiv 0.  \label{r23}
\end{equation}

The construction of the BRST symmetry for the free theory under
consideration starts with the identification of the BRST algebra
on which the BRST differential $s$ acts. The generators of the
BRST algebra are of two kinds: fields/ghosts and antifields. The
ghost spectrum for the model under study comprises the fermionic
ghosts $\eta _{\alpha \beta |\mu }$ associated with the gauge
parameters $\epsilon _{\alpha \beta |\mu }$, as well as the
bosonic ghosts for ghosts $C_{\mu \nu }$ due to the first-stage
reducibility parameters $\theta _{\mu \nu }$. In order to make
compatible the behavior of $\epsilon _{\alpha \beta |\mu }$ and
$\theta _{\mu \nu }$ with that of the corresponding ghosts, we
impose the properties
\begin{equation}
\eta _{\mu \nu |\alpha }=-\eta _{\nu \mu |\alpha },\;\eta _{\left[ \mu \nu
|\alpha \right] }\equiv 0,  \label{r41}
\end{equation}
\begin{equation}
C_{\mu \nu }=-C_{\nu \mu }.  \label{r42}
\end{equation}
The antifield spectrum is organized into the antifields $t^{*\mu \nu |\alpha
\beta }$ of the original tensor field and those of the ghosts, $\eta ^{*\mu
\nu |\alpha }$ and $C^{*\mu \nu }$, of statistics opposite to that of the
associated fields/ghosts. It is understood that the antifields inherit the
mixed symmetry properties of the corresponding fields/ghosts, namely
\begin{equation}
t^{*\mu \nu |\alpha \beta }=-t^{*\nu \mu |\alpha \beta }=-t^{*\mu \nu |\beta
\alpha }=t^{*\alpha \beta |\mu \nu },\;t^{*\left[ \mu \nu |\alpha \right]
\beta }\equiv 0,  \label{r43}
\end{equation}
\begin{equation}
\eta ^{*\mu \nu |\alpha }=-\eta ^{*\nu \mu |\alpha },\;\eta ^{*\left[ \mu
\nu |\alpha \right] }\equiv 0,\;C^{*\mu \nu }=-C^{*\nu \mu }.  \label{r44}
\end{equation}
We will denote the simple and double traces of $t^{*\mu \nu |\alpha \beta }$
by
\begin{equation}
t^{*\nu \beta }=\sigma _{\mu \alpha }t^{*\mu \nu |\alpha \beta },\;t^{*\nu
\beta }=t^{*\beta \nu },\;t^{*}=\sigma _{\nu \beta }t^{*\nu \beta }.
\label{r44a}
\end{equation}

As both the gauge generators and reducibility functions for this model are
field-independent, it follows that the associated BRST differential ($%
s^{2}=0 $) splits into
\begin{equation}
s=\delta +\gamma ,  \label{r45}
\end{equation}
where $\delta $ represents the Koszul-Tate differential ($\delta ^{2}=0$),
graded by the antighost number $\mathrm{agh}$ ($\mathrm{agh}\left( \delta
\right) =-1$), while $\gamma $ stands for the exterior derivative along the
gauge orbits and turns out to be a true differential ($\gamma ^{2}=0$) that
anticommutes with $\delta $ ($\delta \gamma +\gamma \delta =0$), whose
degree is named pure ghost number $\mathrm{pgh}$ ($\mathrm{pgh}\left( \gamma
\right) =1$). These two degrees do not interfere ($\mathrm{agh}\left( \gamma
\right) =0$, $\mathrm{pgh}\left( \delta \right) =0$). The overall degree
that grades the BRST differential is known as the ghost number ($\mathrm{gh}$%
) and is defined like the difference between the pure ghost number and the
antighost number, such that $\mathrm{gh}\left( s\right) =\mathrm{gh}\left(
\delta \right) =\mathrm{gh}\left( \gamma \right) =1$. According to the
standard rules of the BRST method, the corresponding degrees of the
generators from the BRST complex are valued like
\begin{eqnarray}
\mathrm{pgh}\left( t_{\mu \nu |\alpha \beta }\right) &=&0,\;\mathrm{pgh}%
\left( \eta _{\mu \nu |\alpha }\right) =1,\;\mathrm{pgh}\left( C_{\mu \nu
}\right) =2,  \label{r46} \\
\mathrm{pgh}\left( t^{*\mu \nu |\alpha \beta }\right) &=&\mathrm{pgh}\left(
\eta ^{*\mu \nu |\alpha }\right) =\mathrm{pgh}\left( C^{*\mu \nu }\right) =0,
\label{r47} \\
\mathrm{agh}\left( t_{\mu \nu |\alpha \beta }\right) &=&\mathrm{agh}\left(
\eta _{\mu \nu |\alpha }\right) =\mathrm{agh}\left( C_{\mu \nu }\right) =0,
\label{r48} \\
\mathrm{agh}\left( t^{*\mu \nu |\alpha \beta }\right) &=&1,\;\mathrm{agh}%
\left( \eta ^{*\mu \nu |\alpha }\right) =2,\;\mathrm{agh}\left( C^{*\mu \nu
}\right) =3  \label{r49}
\end{eqnarray}
and the actions of $\delta $ and $\gamma $ on them are given by
\begin{equation}
\gamma t_{\mu \nu |\alpha \beta }=\partial _{\mu }\eta _{\alpha \beta |\nu
}-\partial _{\nu }\eta _{\alpha \beta |\mu }+\partial _{\alpha }\eta _{\mu
\nu |\beta }-\partial _{\beta }\eta _{\mu \nu |\alpha },  \label{r50}
\end{equation}
\begin{equation}
\gamma \eta _{\mu \nu |\alpha }=2\partial _{\alpha }C_{\mu \nu }-\partial
_{\left[ \mu \right. }C_{\left. \nu \right] \alpha },\;\gamma C_{\mu \nu }=0,
\label{r51}
\end{equation}
\begin{equation}
\gamma t^{*\mu \nu |\alpha \beta }=0,\;\gamma \eta ^{*\mu \nu |\alpha
}=0,\;\gamma C^{*\mu \nu }=0,  \label{r54}
\end{equation}
\begin{equation}
\delta t_{\mu \nu |\alpha \beta }=0,\;\delta \eta _{\mu \nu |\alpha
}=0,\;\delta C_{\mu \nu }=0,  \label{r52}
\end{equation}
\begin{equation}
\delta t^{*\mu \nu |\alpha \beta }=\frac{1}{4}T^{\mu \nu |\alpha \beta
},\;\delta \eta ^{*\alpha \beta |\nu }=-4\partial _{\mu }t^{*\mu \nu |\alpha
\beta },\;\delta C^{*\mu \nu }=3\partial _{\alpha }\eta ^{*\mu \nu |\alpha },
\label{r53}
\end{equation}
with $T_{\mu \nu |\alpha \beta }$ introduced in (\ref{r15}) and both $\delta
$ and $\gamma $ taken to act like right derivations.

\section{Cohomology of $\gamma $ and related matters\label{hgama}}

The main aim of this paper is to study of the local cohomology $H\left(
s|d\right) $ in form degree $D$ ($D\geq 5$). As it will be further seen, an
indispensable ingredient in the computation of $H\left( s|d\right) $ is the
cohomology algebra of the exterior longitudinal derivative ($H\left( \gamma
\right) $). It is defined by the equivalence classes of $\gamma $-closed
non-integrated densities $a$ of fields, ghosts, antifields and their
spacetime derivatives, $\gamma a=0$, modulo $\gamma $-exact terms. If $a\in
H\left( \gamma \right) $ is $\gamma $-exact, $a=\gamma b$, then $a$ belongs
to the class of the element zero and we call it $\gamma $-trivial. In other
words, the solution to the equation $\gamma a=0$ is unique up to $\gamma $%
-trivial objects, $a\rightarrow a+\gamma b$. The cohomology algebra $H\left(
\gamma \right) $ inherits a natural grading $H\left( \gamma \right)
=\bigoplus\nolimits_{l\geq 0}H^{l}\left( \gamma \right) $, where $l$ is the
pure ghost number. Let $a$ be an element of $H\left( \gamma \right) $ with
definite pure ghost number, antighost number and form degree ($\deg $)
\begin{equation}
\gamma a=0,\;\mathrm{pgh}\left( a\right) =l\geq 0,\;\mathrm{agh}\left(
a\right) =k\geq 0,\;\deg \left( a\right) =p\leq D.  \label{r67}
\end{equation}
In the sequel we analyze the general form of $a$ with the above properties
with the help of the definitions (\ref{r50}--\ref{r54}).

The formula (\ref{r54}) shows that all the antifields
\begin{equation}
\chi ^{*\Delta }=\left( t^{*\mu \nu |\alpha \beta },\eta ^{*\mu \nu |\alpha
},C^{*\mu \nu }\right)   \label{r73ab}
\end{equation}
belong (non-trivially) to $H^{0}\left( \gamma \right) $. From the definition
(\ref{r50}) we infer that the most general $\gamma $-closed (and obviously
non-trivial) elements constructed in terms of the original tensor field are
the components of the curvature tensor (\ref{curv}) and their spacetime
derivatives, so all these pertain to $H^{0}\left( \gamma \right) $. Using
the first definition in (\ref{r51}), we notice that there is no $\gamma $%
-closed linear combination of the undifferentiated ghosts of pure ghost
number one. After some computation, we find that the most general $\gamma $
-closed quantities in the first-order derivatives of the pure ghost number
one ghosts have the mixed symmetry of the tensor field $t_{\mu \nu |\alpha
\beta }$ itself
\begin{equation}
M_{\mu \nu |\alpha \beta }=\partial _{\mu }\eta _{\alpha \beta |\nu
}-\partial _{\nu }\eta _{\alpha \beta |\mu }+\partial _{\alpha }\eta _{\mu
\nu |\beta }-\partial _{\beta }\eta _{\mu \nu |\alpha }.  \label{r75a}
\end{equation}
It is easy to see from formula (\ref{r50}) that $M_{\mu \nu |\alpha \beta }$
is $\gamma $-exact
\begin{equation}
M_{\mu \nu |\alpha \beta }=\gamma t_{\mu \nu |\alpha \beta },  \label{r76}
\end{equation}
and thus it must be discarded from $H^{1}\left( \gamma \right) $ as being
trivial. Along the same line, one can prove that the only $\gamma $-closed
combinations with $N\geq 2$ spacetime derivatives of the ghosts $\eta _{\mu
\nu |\alpha }$ are actually polynomials with $\left( N-1\right) $
derivatives in the elements $M_{\mu \nu |\alpha \beta }$, which, by means of
(\ref{r76}), are $\gamma $-exact, and hence trivial in $H^{1}\left( \gamma
\right) $. In conclusion, there is no non-trivial object constructed out of
the ghosts $\eta _{\mu \nu |\alpha }$ and their derivatives in $H^{1}\left(
\gamma \right) $, which implies that $H^{1}\left( \gamma \right) =0$ as
there are no other ghosts of pure ghost number equal to one in the BRST
complex. The BRST complex for the model under consideration contains no
other ghosts with odd pure ghost numbers, so we conclude that
\begin{equation}
H^{2l+1}\left( \gamma \right) =0,\;\mathrm{for}\;\mathrm{all}\;l\geq 0.
\label{hgzero}
\end{equation}
The definitions (\ref{r51}) show that the undifferentiated ghosts of pure
ghost number equal to two, $C_{\mu \nu }$, belong to $H\left( \gamma \right)
$. The $\gamma $- closedness of $C_{\mu \nu }$ further implies that all
their derivatives are also $\gamma $-closed. Regarding their first-order
derivatives, from the first relation in (\ref{r51}) we observe that their
symmetric part is $\gamma $-exact
\begin{equation}
\partial _{\left( \mu \right. }C_{\left. \nu \right) \alpha }\equiv \gamma
\left( -\frac{1}{3}\eta _{\alpha \left( \mu |\nu \right) }\right) ,
\label{r77}
\end{equation}
where $\left( \mu \nu \cdots \right) $ denotes plain symmetrization with
respect to the indices between brackets without normalization factors, such
that $\partial _{\left( \mu \right. }C_{\left. \nu \right) \alpha }$ will be
removed from $H\left( \gamma \right) $. Meanwhile, their antisymmetric part $%
\partial _{\left[ \mu \right. }C_{\left. \nu \right] \alpha }$ is not $%
\gamma $-exact, and hence can be taken as a non-trivial representative of $%
H\left( \gamma \right) $. After some calculations, we find that all the
second-order derivatives of the ghosts for ghosts are $\gamma $-exact
\begin{equation}
\partial _{\alpha }\partial _{\beta }C_{\mu \nu }=\frac{1}{12}\gamma \left(
3\left( \partial _{\alpha }\eta _{\mu \nu |\beta }+\partial _{\beta }\eta
_{\mu \nu |\alpha }\right) +\partial _{\left[ \mu \right. }\eta _{\left. \nu
\right] \,\left( \alpha |\beta \right) }\right) ,  \label{r78}
\end{equation}
and so will be their higher-order derivatives. In conclusion, the only
non-trivial combinations in $H\left( \gamma \right) $ constructed from the
ghosts of pure ghost number equal to two are polynomials in $C_{\mu \nu }$
and $\partial _{\left[ \mu \right. }C_{\left. \nu \right] \alpha }$.
Combining this result with the previous one on $H^{0}\left( \gamma \right) $
being non-vanishing, we have actually proved that only the even
cohomological spaces of the exterior longitudinal derivative, $H^{2l}\left(
\gamma \right) $ with $l\geq 0$, are non-vanishing.

According to the results exposed so far, we can state that the \emph{general
local solution} to the equation (\ref{r67}) for $\mathrm{pgh}\left( a\right)
=2l>0$ is, up to trivial, $\gamma $-exact contributions, of the type
\begin{equation}
a=\sum_{J}\alpha _{J}\left( \left[ \chi ^{*\Delta }\right] ,\left[ F_{\mu
\nu \lambda |\alpha \beta \gamma }\right] \right) e^{J}\left( C_{\mu \nu
},\partial _{\left[ \mu \right. }C_{\left. \nu \right] \alpha }\right) ,
\label{r79}
\end{equation}
where the notation $f\left( \left[ q\right] \right) $ means that the
function $f$ depends on the variable $q$ and its subsequent derivatives up
to a finite number. In the above, $e^{J}$ are the elements of pure ghost
number $2l$ (and obviously of antighost number zero) of a basis in the space
of polynomials in $C_{\mu \nu }$ and $\partial _{\left[ \mu \right.
}C_{\left. \nu \right] \alpha }$
\begin{equation}
\mathrm{pgh}\left( e^{J}\right) =2l>0,\;\mathrm{agh}\left( e^{J}\right) =0.
\label{t82}
\end{equation}
The objects $\alpha _{J}$ (obviously non-trivial in $H^{0}\left( \gamma
\right) $) were taken to have a bounded number of derivatives, and therefore
they are polynomials in the antifields $\chi ^{*\Delta }$, in the curvature
tensor $F_{\mu \nu \lambda |\alpha \beta \gamma }$, as well as in their
derivatives. In agreement with (\ref{r67}), they display the properties
\begin{equation}
\mathrm{pgh}\left( \alpha _{J}\right) =0,\;\mathrm{agh}\left( \alpha
_{J}\right) =k\geq 0,\;\deg \left( \alpha _{J}\right) =p\leq D.  \label{t83}
\end{equation}
In the case $l=0$, the general (non-trivial) local elements of $H\left(
\gamma \right) $ are precisely $\alpha _{J}\left( \left[ \chi ^{*\Delta
}\right] ,\left[ F_{\mu \nu \lambda |\alpha \beta \gamma }\right] \right) $,
which will be called ``invariant polynomials'' in what follows. At zero
antighost number, the invariant polynomials are polynomials in the curvature
tensor $F_{\mu \nu \lambda |\alpha \beta \gamma }$ and its derivatives.

In order to analyze the local cohomology $H\left( s|d\right) $ we are going
to need, besides $H\left( \gamma \right) $, also the cohomology of the
exterior spacetime differential $H\left( d\right) $ in the space of
invariant polynomials and other basic properties, which are addressed below.

\begin{theorem}
\label{hginv}The cohomology of $d$ in form degree strictly less than $D$ is
trivial in the space of invariant polynomials with strictly positive
antighost number. This means that the conditions
\begin{equation}
\gamma \alpha =0,\;d\alpha =0,\;\mathrm{agh}\left( \alpha \right) >0,\;\deg
\alpha <D,\;\alpha =\alpha \left( \left[ \chi ^{*\Delta }\right] ,\left[
F\right] \right) ,  \label{A1}
\end{equation}
imply
\begin{equation}
\alpha =d\beta ,  \label{A2}
\end{equation}
for some invariant polynomial $\beta \left( \left[ \chi ^{*\Delta }\right]
,\left[ F\right] \right) $.
\end{theorem}

\emph{Proof} In (\ref{A1}), the notation $F$ signifies the curvature tensor,
of components $F_{\mu \nu \lambda |\alpha \beta \gamma }$, and $\chi
^{*\Delta }$ is explained in (\ref{r73ab}). Meanwhile, $\deg \alpha $ is the
form degree of $\alpha $. In order to prove the theorem, we decompose $d$ as
\begin{equation}
d=d_{0}+d_{1},  \label{A4}
\end{equation}
where $d_{1}$ acts on the antifields $\chi ^{*\Delta }$ and their
derivatives only, while $d_{0}$ acts on the curvature tensor and its
derivatives
\begin{equation}
d_{0}=\partial _{\mu _{1}}^{0}dx^{\mu _{1}},\;d_{1}=\partial _{\mu
_{1}}^{1}dx^{\mu _{1}},  \label{A5}
\end{equation}
with
\begin{equation}
\partial _{\mu _{1}}^{0}=F_{\mu \nu \lambda |\alpha \beta \gamma ,\mu _{1}}%
\frac{\partial }{\partial F_{\mu \nu \lambda |\alpha \beta \gamma }}+F_{\mu
\nu \lambda |\alpha \beta \gamma ,\mu _{1}\mu _{2}}\frac{\partial }{\partial
F_{\mu \nu \lambda |\alpha \beta \gamma ,\mu _{2}}}+\cdots ,  \label{A6}
\end{equation}
\begin{equation}
\partial _{\mu _{1}}^{1}=\chi _{\;\;\;\;,\mu _{1}}^{*\Delta }\frac{\partial
^{L}}{\partial \chi ^{*\Delta }}+\chi _{\;\;\;\;,\mu _{1}\mu _{2}}^{*\Delta }%
\frac{\partial ^{L}}{\partial \chi _{\;\;\;\;,\mu _{2}}^{*\Delta }}+\cdots .
\label{A7}
\end{equation}
We used the common convention $f_{,\mu _{1}}\equiv \partial f/\partial
x^{\mu _{1}}$. Obviously, $d^{2}=0$ on invariant polynomials is equivalent
with the nilpotency and anticommutation of its components acting on
invariant polynomials
\begin{equation}
d_{0}^{2}=0=d_{1}^{2},\;d_{0}d_{1}+d_{1}d_{0}=0.  \label{A8}
\end{equation}
The action of $d_{0}$ on a given invariant polynomial with say $l$
derivatives of $F$ and $j$ derivatives of $\chi ^{*\Delta }$ results in an
invariant polynomial with $\left( l+1\right) $ derivatives of $F$ and $j$
derivatives of $\chi ^{*\Delta }$, while the action of $d_{1}$ on the same
object leads to an invariant polynomial with $l$ derivatives of $F$ and $%
\left( j+1\right) $ derivatives of $\chi ^{*\Delta }$. In particular, $d_{0}$
gives zero when acting on an invariant polynomial that does not involve the
curvature or its derivatives, and the same is valid with respect to $d_{1}$
acting on an invariant polynomial that does not depend on any of the
antifields or their derivatives. From (\ref{A6}--\ref{A7}) we observe that
\begin{equation}
\mathrm{agh}\left( d_{0}\right) =\mathrm{agh}\left( d_{1}\right) =\mathrm{agh%
}\left( d\right) =0,  \label{A8a}
\end{equation}
such that neither of them change the antighost number of the objects on
which they act.

The antifields $\chi ^{*\Delta }$ verify no relations between themselves and
their derivatives, except the usual symmetry properties of the type $\chi
_{\;\;\;\;,\mu _{1}\mu _{2}}^{*\Delta }=\chi _{\;\;\;\;,\mu _{2}\mu
_{1}}^{*\Delta }$, and accordingly will be named ``foreground'' fields. On
the contrary, the derivatives of the components of the curvature tensor
satisfy the Bianchi II identities (\ref{r23}), and in view of this we say
that $F_{\mu \nu \lambda |\alpha \beta \gamma }$ are ``background'' fields.
So, $d_{0}$ acts only on the background fields and their derivatives, while $%
d_{1}$ acts only on the foreground fields and their derivatives. According
to the proposition on page 363 in \cite{dubplb}, we have that the entire
cohomology of $d_{1}$ in form degree strictly less than $D$ is trivial in
the space of invariant polynomials with strictly positive antighost number.
This means that
\begin{equation}
\alpha =\alpha \left( \left[ \chi ^{*\Delta }\right] ,\left[ F\right]
\right) ,\;\mathrm{agh}\left( \alpha \right) =k>0,\;\deg \left( \alpha
\right) =p<D,\;d_{1}\alpha =0,  \label{A9}
\end{equation}
implies that
\begin{equation}
\alpha =d_{1}\beta ,  \label{A10}
\end{equation}
with
\begin{equation}
\beta =\beta \left( \left[ \chi ^{*\Delta }\right] ,\left[ F\right] \right)
,\;\mathrm{agh}\left( \beta \right) =k>0,\;\deg \left( \beta \right) =p-1.
\label{A11}
\end{equation}
In particular, we have that if an invariant polynomial (of form degree $p<D$
and with strictly positive antighost number) depending only on the
undifferentiated antifields is $d_{1}$-closed, then it vanishes
\begin{eqnarray}
&&\left( \bar{\alpha}=\bar{\alpha}\left( \chi ^{*\Delta },\left[ F\right]
\right) ,\;\mathrm{agh}\left( \bar{\alpha}\right) >0,\right.  \nonumber \\
&&\left. \deg \left( \bar{\alpha}\right) =p<D,\;d_{1}\bar{\alpha}=0\right)
\Rightarrow \bar{\alpha}=0.  \label{A11a}
\end{eqnarray}
Only $d_{0}$ has non-trivial cohomology. For instance, any form depending
only on the antifields and their derivatives is $d_{0}$-closed, but it is
clearly not $d_{0}$-exact.

Next, assume that $\alpha $ is a homogeneous form of degree $p<D$ and
antighost number $k>0$ that satisfies the conditions (\ref{A1}). We
decompose $\alpha $ according to the number of derivatives of the antifields
\begin{equation}
\alpha =\stackrel{(0)}{\alpha }+\stackrel{(1)}{\alpha }+\cdots +\stackrel{(s)%
}{\alpha },\;\mathrm{agh}\left( \stackrel{(i)}{\alpha }\right) =k>0,\;\deg
\left( \stackrel{(i)}{\alpha }\right) =p<D,  \label{A12}
\end{equation}
where $\stackrel{(i)}{\alpha }$ signifies the component from $\alpha $ with $%
i$ derivatives of the antifields. (The decomposition contains a finite
number of terms since $\alpha $ is local by assumption.) As $\alpha $ is an
invariant polynomial of form degree $p<D$ and strictly positive antighost
number, each component $\left( \stackrel{(i)}{\alpha }\right) _{0\leq i\leq
s}$ is an invariant polynomial with the same form degree and strictly
positive antighost number. The proof of the theorem is realized in $\left(
s+1\right) $ steps.

Step 1. Taking into account the splitting (\ref{A4}), the projection of the
equation
\begin{equation}
d\alpha =0  \label{A12a}
\end{equation}
on the maximum number of derivatives of the antifields ($s+1$) produces
\begin{equation}
d_{1}\stackrel{(s)}{\alpha }=0,  \label{A13}
\end{equation}
and hence the triviality of the cohomology of $d_{1}$ ensures that
\begin{equation}
\stackrel{(s)}{\alpha }=d_{1}\stackrel{(s-1)}{\beta },\;\mathrm{agh}\left(
\stackrel{(s-1)}{\beta }\right) =k>0,\;\deg \left( \stackrel{(s-1)}{\beta }%
\right) =p-1,  \label{A14}
\end{equation}
where $\stackrel{(s-1)}{\beta }$ is an invariant polynomial of form degree ($%
p-1$), with strictly positive antighost number and containing only ($s-1$)
derivatives of the antifields. If we introduce the $p$-form
\begin{equation}
\alpha _{1}=\alpha -d\stackrel{(s-1)}{\beta },  \label{A15}
\end{equation}
then the equation (\ref{A12a}) together with the nilpotency of $d$ further
yield
\begin{equation}
d\alpha _{1}=0.  \label{A16}
\end{equation}
It is by construction an invariant polynomial of form degree $p$ and of
strictly positive antighost number and, most important, the maximum number
of derivatives of the antifields from $\alpha _{1}$ is equal to ($s-1$).
Indeed, if we replace (\ref{A14}) in (\ref{A12}) and then in (\ref{A15}), we
get that
\begin{equation}
\alpha _{1}=\stackrel{(0)}{\alpha }+\stackrel{(1)}{\alpha }+\cdots +%
\stackrel{(s-2)}{\alpha }+\stackrel{(s-1)}{\alpha }-d_{0}\stackrel{(s-1)}{%
\beta }.  \label{A17}
\end{equation}
Then, the maximum number of derivatives of the antifields from the first $s$
terms in the right-hand side of (\ref{A17}) is contained in $\stackrel{(s-1)%
}{\alpha }$, being equal to ($s-1$), while $d_{0}\stackrel{(s-1)}{\beta }$
has the same number of derivatives of the antifields like $\stackrel{(s-1)}{%
\beta }$, which is again ($s-1$).

Step 2. If we project now the equation (\ref{A16}) on the maximum number of
derivatives of the antifields ($s$), we infer that
\begin{equation}
d_{1}\left( \stackrel{(s-1)}{\alpha }-d_{0}\stackrel{(s-1)}{\beta }\right)
=0,  \label{A18}
\end{equation}
with $\stackrel{(s-1)}{\alpha }-d_{0}\stackrel{(s-1)}{\beta }$ an invariant
polynomial of form degree $p$ and of strictly positive antighost number.
Using again the triviality of the cohomology of $d_{1}$, we deduce that
\begin{equation}
\stackrel{(s-1)}{\alpha }-d_{0}\stackrel{(s-1)}{\beta }=d_{1}\stackrel{(s-2)%
}{\beta },\;\mathrm{agh}\left( \stackrel{(s-2)}{\beta }\right) =k>0,\;\deg
\left( \stackrel{(s-2)}{\beta }\right) =p-1,  \label{A19}
\end{equation}
where $\stackrel{(s-2)}{\beta }$ is an invariant polynomial of form degree ($%
p-1$), with strictly positive antighost number and containing only ($s-2$)
derivatives of the antifields. At this stage, we define the $p$-form
\begin{equation}
\alpha _{2}=\alpha -d\left( \stackrel{(s-1)}{\beta }+\stackrel{(s-2)}{\beta }%
\right) .  \label{A20}
\end{equation}
The equation (\ref{A12a}) together with the nilpotency of $d$ further yield
\begin{equation}
d\alpha _{2}=0.  \label{A21}
\end{equation}
Clearly, $\alpha _{2}$ is an invariant polynomial of form degree $p$ and of
strictly positive antighost number. It is essential to remark that the
maximum number of derivatives of the antifields from $\alpha _{2}$ is equal
to ($s-2$). This results by inserting (\ref{A14}) and (\ref{A19}) in (\ref
{A12}) and consequently in (\ref{A20}), which then gives
\begin{equation}
\alpha _{2}=\stackrel{(0)}{\alpha }+\stackrel{(1)}{\alpha }+\cdots +%
\stackrel{(s-3)}{\alpha }+\stackrel{(s-2)}{\alpha }-d_{0}\stackrel{(s-2)}{%
\beta }.  \label{A22}
\end{equation}
etc.

Step $s$. Proceeding in the same manner, at the $s$-th step we obtain an
invariant polynomial of form degree $p$ and with strictly positive antighost
number, which contains only the undifferentiated antifields
\begin{eqnarray}
&&\alpha _{s}=\alpha -d\left( \stackrel{(s-1)}{\beta }+\cdots +\stackrel{(0)%
}{\beta }\right) =\stackrel{(0)}{\alpha }-d_{0}\stackrel{(0)}{\beta },
\label{A23} \\
&&\mathrm{agh}\left( \stackrel{(j)}{\beta }\right) =k>0,\;\deg \left(
\stackrel{(j)}{\beta }\right) =p-1,\;0\leq j\leq s-1.  \label{A23a}
\end{eqnarray}
(All $\left( \stackrel{(j)}{\beta }\right) 0\leq j\leq s-1$ are invariant
polynomials.) The equation (\ref{A12a}) and the nilpotency of $d$ lead to
the equation
\begin{equation}
d\alpha _{s}=0.  \label{A24}
\end{equation}

Step ($s+1$). The projection of (\ref{A24}) on the maximum number of
derivatives of the antifields (one) is
\begin{equation}
d_{1}\left( \stackrel{(0)}{\alpha }-d_{0}\stackrel{(0)}{\beta }\right) =0,\;%
\mathrm{agh}\left( \stackrel{(0)}{\alpha }-d_{0}\stackrel{(0)}{\beta }%
\right) =k>0.  \label{A25}
\end{equation}
From (\ref{A25}) and (\ref{A11a}) (with $\bar{\alpha}$ replaced by $%
\stackrel{(0)}{\alpha }-d_{0}\stackrel{(0)}{\beta }$) we get that
\begin{equation}
\stackrel{(0)}{\alpha }-d_{0}\stackrel{(0)}{\beta }=0,  \label{A26}
\end{equation}
which substituted in (\ref{A23}) finally allows us to write that
\begin{equation}
\alpha =d\beta ,  \label{A27}
\end{equation}
with
\begin{equation}
\beta =\left( \stackrel{(s-1)}{\beta }+\cdots +\stackrel{(0)}{\beta }\right)
,\;\mathrm{agh}\left( \beta \right) =k>0,\;\deg \left( \beta \right) =p-1,
\label{A28}
\end{equation}
and this proves the theorem since $\beta $ is an invariant polynomial of
form degree ($p-1$) and with strictly positive antighost number. $%
\blacksquare $

In form degree $D$ the Theorem \ref{hginv} is replaced with: let $\alpha
=\rho dx^{0}\wedge \cdots \wedge dx^{D-1}$ be a $d$-exact invariant
polynomial of form degree $D$ and of strictly positive antighost number, $%
\mathrm{agh}\left( \alpha \right) =k>0$, $\mathrm{\deg }\left( \alpha
\right) =D$, $\alpha =d\beta $. Then, one can take the $\left( D-1\right) $%
-form $\beta $ to be an invariant polynomial (of antighost number $k$). In
dual notations, this means that if $\rho $ with $\mathrm{agh}\left( \rho
\right) =k>0$ is an invariant polynomial whose Euler-Lagrange derivatives
are all vanishing, $\rho =\partial _{\mu }j^{\mu }$, then $j^{\mu }$ can be
taken to be also invariant. Theorem \ref{hginv} can be generalized as
follows.

\begin{theorem}
\label{hdinhg}The cohomology of $d$ computed in $H\left( \gamma \right) $ is
trivial in form degree strictly less than $D$ and in strictly positive
antighost number
\begin{equation}
H_{p}^{g,k}\left( d,H\left( \gamma \right) \right) =0,\;k>0,\;p<D,
\label{A29}
\end{equation}
where $p$ is the form degree, $k$ is the antighost number and $g$ is the
ghost number.
\end{theorem}

\emph{Proof} An element $a$ from $H_{p}^{g,k}\left( d,H\left( \gamma \right)
\right) $ is a $p$-form of definite ghost number $g$ and antighost number $k$%
, pertaining to the cohomology of $\gamma $, which is $d$-closed modulo $%
\gamma $%
\begin{equation}
\gamma a=0,\;da=\gamma \mu ,\;\mathrm{agh}\left( a\right) =k,\;\mathrm{gh}%
\left( a\right) =g,\;\deg \left( a\right) =p.  \label{A30}
\end{equation}
The theorem states that if $a$ satisfies the conditions (\ref{A30}) with $%
p<D $ and $k>0$, then $a$ is trivial in $H_{p}^{g,k}\left( d,H\left( \gamma
\right) \right) $%
\begin{equation}
a=d\nu +\gamma \rho ,\;\gamma \nu =0,  \label{A31}
\end{equation}
where
\begin{eqnarray}
\mathrm{agh}\left( \nu \right) &=&\mathrm{agh}\left( \rho \right) =k>0,\;%
\mathrm{gh}\left( \nu \right) =g,\;\mathrm{gh}\left( \rho \right) =g-1,
\label{A31a} \\
\deg \left( \nu \right) &=&p-1,\;\deg \left( \rho \right) =p<D.  \label{A31b}
\end{eqnarray}
Since $g=l^{\prime }-k$, with $l^{\prime }$ the pure ghost number of $a$,
and $l^{\prime }$ takes positive values $l^{\prime }\geq 0$, it follows that
$g$ is restricted to fulfill the condition $g\geq -k$. Thus, if $g<-k$, then
$a=0$. The theorem is thus trivially obeyed for $g<-k$. Also, due to the
fact that $H^{2l+1}\left( \gamma \right) =0$ for all $l\geq 0$, we find
again that for $g=2l+1-k$, with $l\geq 0$, the theorem is fulfilled. In the
case $g=-k$ we have that $\mathrm{pgh}\left( a\right) =0$, and therefore $a$
depends only on the antifields, on the field $t_{\mu \nu |\alpha \beta }$
and their derivatives. The $\gamma $-closedness of $a$, $\gamma a=0$, then
induces that $a$ is actually an invariant polynomial of strictly positive
antighost number and of form degree strictly less than $D$, and hence
Theorem \ref{hdinhg} reduces to the Theorem \ref{hginv}, which has been
proved in the above. Thus, for $g=-k$, (\ref{A30}) and (\ref{A31}) must be
replaced with (\ref{A1}), respectively (\ref{A2}) (or, in other words, we
must set $\mu =0$ in (\ref{A30}) and $\rho =0$ in (\ref{A31})).
Consequently, we only need to prove the theorem for $g=2l-k$ and $l>0$. This
is done below.

We consider a non-trivial element $a$ from $H\left( \gamma \right) $ of form
degree $p<D$, of antighost number $k>0$ and of ghost number $g=2l-k$, for $%
l>0$
\begin{equation}
\gamma a=0,\;\mathrm{agh}\left( a\right) =k>0,\;\mathrm{gh}\left( a\right)
=g=2l-k,\;\deg \left( a\right) =p<D,\;l>0.  \label{A33}
\end{equation}
According to the results from the Subsection \ref{hgama}, $a$ has the form
(up to trivial, $\gamma $-exact contributions)
\begin{equation}
a=\sum \alpha _{J}\left( \left[ \chi ^{*\Delta }\right] ,\left[ F\right]
\right) e^{J}\left( C_{\mu \nu },\partial _{\left[ \mu \right. }C_{\left.
\nu \right] \alpha }\right) ,  \label{A34}
\end{equation}
where $\alpha _{J}\left( \left[ \chi ^{*\Delta }\right] ,\left[ F\right]
\right) $ are invariant polynomials of antighost number $k$ and form degree $%
p$, while $e^{J}\left( C_{\mu \nu },\partial _{\left[ \mu \right. }C_{\left.
\nu \right] \alpha }\right) $ represent the elements of pure ghost number $%
2l>0$ of a basis in the space of polynomials in $C_{\mu \nu }$ and $\partial
_{\left[ \mu \right. }C_{\left. \nu \right] \alpha }$ (there are precisely $%
\left( l+1\right) $ elements $e^{J}$ of pure ghost number $2l$, which
commute among themselves)
\begin{equation}
\mathrm{agh}\left( \alpha _{J}\right) =k>0,\;\deg \left( \alpha _{J}\right)
=p<D,\;\mathrm{pgh}\left( e^{J}\right) =2l>0,\;\mathrm{for\;all}\;J.
\label{A35}
\end{equation}
We will use in extenso the following obvious properties
\begin{equation}
\gamma ^{2}=0,\;d^{2}=0,\;\gamma d+d\gamma =0,\;\mathrm{pgh}\left( d\right)
=0,\;\deg \left( \gamma \right) =0,  \label{A35a}
\end{equation}
\begin{equation}
\sum \alpha _{J}\left( \left[ \chi ^{*\Delta }\right] ,\left[ F\right]
\right) e^{J}\left( C_{\mu \nu },\partial _{\left[ \mu \right. }C_{\left.
\nu \right] \alpha }\right) =\gamma \left( \mathrm{something}\right)
\Leftrightarrow \alpha _{J}=0,\;\mathrm{for\;all\;}J,  \label{A36}
\end{equation}
\begin{equation}
d\alpha _{J}\left( \left[ \chi ^{*\Delta }\right] ,\left[ F\right] \right)
=\alpha _{J}^{\prime }\left( \left[ \chi ^{*\Delta }\right] ,\left[ F\right]
\right) ,  \label{A38}
\end{equation}
where
\begin{equation}
\mathrm{agh}\left( \alpha _{J}^{\prime }\right) =\mathrm{agh}\left( \alpha
_{J}\right) ,\;\deg \left( \alpha _{J}^{\prime }\right) =\deg \left( \alpha
_{J}\right) +1.  \label{A40}
\end{equation}
It is useful to define an operator $\bar{D}$ that acts only on $H\left(
\gamma \right) $ via the relations
\begin{eqnarray}
\bar{D}\alpha \left( \left[ \chi ^{*\Delta }\right] ,\left[ F\right] \right)
&=&d\alpha \left( \left[ \chi ^{*\Delta }\right] ,\left[ F\right] \right) ,
\label{A40A} \\
\bar{D}C_{\mu \nu } &=&\frac{1}{2}\partial _{\left[ \alpha \right.
}C_{\left. \mu \right] \nu }dx^{\alpha },  \label{A40B} \\
\bar{D}\partial _{\left[ \alpha \right. }C_{\left. \mu \right] \nu } &=&0,
\label{A40C} \\
\bar{D}\left( \gamma b\right) &=&0,  \label{A40CA}
\end{eqnarray}
which is easily seen to be a differential in $H\left( \gamma \right) $, $%
\bar{D}^{2}a=0$ for any $a$ with $\gamma a=0$. According to the relation (%
\ref{r77}), we have that
\begin{eqnarray}
dC_{\mu \nu } &=&\left( \partial _{\alpha }C_{\mu \nu }\right) dx^{\alpha }=%
\frac{1}{2}\partial _{\left[ \alpha \right. }C_{\left. \mu \right] \nu
}dx^{\alpha }+\frac{1}{2}\partial _{\left( \alpha \right. }C_{\left. \mu
\right) \nu }dx^{\alpha }  \nonumber \\
&=&\bar{D}C_{\mu \nu }+\gamma \left( \frac{1}{6}\eta _{\nu \left( \alpha
|\mu \right) }dx^{\alpha }\right) ,  \label{A40D}
\end{eqnarray}
while from (\ref{r78}) and (\ref{A40C}) we find that
\begin{eqnarray}
d\partial _{\left[ \alpha \right. }C_{\left. \mu \right] \nu } &=&\left(
\partial _{\rho }\partial _{\left[ \alpha \right. }C_{\left. \mu \right] \nu
}\right) dx^{\rho }=\bar{D}\partial _{\left[ \alpha \right. }C_{\left. \mu
\right] \nu }  \nonumber \\
&&+\gamma \left( -\frac{1}{12}\left( 3\left( \partial _{\alpha }\eta _{\mu
\nu |\rho }+\partial _{\rho }\eta _{\mu \nu |\alpha }\right) +\partial
_{\left[ \mu \right. }\eta _{\left. \nu \right] \,\left( \rho |\alpha
\right) }\right. \right.  \nonumber \\
&&\left. \left. -3\left( \partial _{\mu }\eta _{\alpha \nu |\rho }+\partial
_{\rho }\eta _{\alpha \nu |\mu }\right) -\partial _{\left[ \alpha \right.
}\eta _{\left. \nu \right] \,\left( \rho |\mu \right) }\right) dx^{\rho
}\right) .  \label{A40E}
\end{eqnarray}
Moreover, from (\ref{A40B}--\ref{A40C}) we observe that
\begin{equation}
\bar{D}e^{J}=A_{I}^{J}e^{I},  \label{A40H}
\end{equation}
for some constant matrix of elements $A_{I}^{J}$, that involves $dx^{\rho }$%
, such that $\bar{D}e^{J}$ is $\gamma $-closed, but not $\gamma $-exact
\begin{equation}
de^{J}=\bar{D}e^{J}+\gamma \hat{e}^{J}=\sum_{I}A_{I}^{J}e^{I}+\gamma \hat{e}%
^{J},  \label{A40I}
\end{equation}
where $\hat{e}^{J}$ depends in general on $C_{\mu \nu }$, $\partial _{\left[
\alpha \right. }C_{\left. \mu \right] \nu }$ and $\left[ \eta _{\mu \nu
|\alpha }\right] $. Taking into account the relations (\ref{A40A}), (\ref
{A40CA}) and (\ref{A40I}), we conclude that the differential $d$ on $H\left(
\gamma \right) $ coincides with $\bar{D}$%
\begin{equation}
da=\bar{D}a+\gamma b,\;a\in H\left( \gamma \right) ,  \label{dDgamma}
\end{equation}
where $\bar{D}a$ either vanishes or is $\gamma $-non-trivial. It is easy to
see that $d$ induces a well defined differential in $H\left( \gamma \right) $%
, which can be taken to be $\bar{D}$. Indeed, suppose that $a$ from $H\left(
\gamma \right) $ is non-trivial and fulfills the properties (\ref{A33}),
such that it can be expressed in the form (\ref{A34}), with $\alpha _{J}$
and $e^{J}$ subject to (\ref{A35}). Accordingly, from (\ref{A40A}) and (\ref
{A40I}) we can write that
\begin{equation}
da=\bar{D}a+\gamma \left( \sum \alpha _{J}\hat{e}^{J}\right) ,
\label{eqnew1}
\end{equation}
where
\begin{equation}
\bar{D}a=\sum \left( \alpha _{J}\bar{D}e^{J}+\left( d\alpha _{J}\right)
e^{J}\right) ,  \label{eqnew2}
\end{equation}
is a non-trivial element from $H\left( \gamma \right) $ due to (\ref{A36}--%
\ref{A38}) and (\ref{A40H}). It follows that if $a$ is non-trivial in $%
H\left( \gamma \right) $, then $da$ also defines a non-trivial element from $%
H\left( \gamma \right) $, which is in the same equivalence class with $\bar{D%
}a$. On the other hand, if $a$ from $H\left( \gamma \right) $ is trivial, $%
a=\gamma b$, then the anticommutation between $d$ and $\gamma $ together
with (\ref{A40CA}) yield that $da=\gamma \left( -db\right) $, and thus $da$
is also trivial in $H\left( \gamma \right) $, belonging to the class of the
element zero, just like $\bar{D}\left( \gamma b\right) $. As a consequence
of the above discussion, we can state that
\begin{equation}
H_{p}^{g,k}\left( d,H\left( \gamma \right) \right) \simeq H_{p}^{g,k}\left(
\bar{D}\right) ,  \label{A41a}
\end{equation}
so in order to prove the theorem it is enough to prove that $%
H_{p}^{g,k}\left( \bar{D}\right) =0$ for $g=2l-k$, with $l,k>0$ and $p<D$.

To this end, we decompose $\bar{D}$ as a sum of two operators
\begin{equation}
\bar{D}=\bar{D}_{0}+\bar{D}_{1},  \label{A42}
\end{equation}
defined through
\begin{eqnarray}
\bar{D}_{0}\alpha \left( \left[ \chi ^{*\Delta }\right] ,\left[ F\right]
\right) &=&\bar{D}\alpha \left( \left[ \chi ^{*\Delta }\right] ,\left[
F\right] \right) =d\alpha ,  \label{A43a} \\
\bar{D}_{0}C_{\mu \nu } &=&0,  \label{A43b} \\
\bar{D}_{0}\partial _{\left[ \alpha \right. }C_{\left. \mu \right] \nu }
&=&0,  \label{A43c} \\
\bar{D}_{0}\left( \gamma b\right) &=&0,  \label{A43d} \\
\bar{D}_{1}\alpha \left( \left[ \chi ^{*\Delta }\right] ,\left[ F\right]
\right) &=&0,  \label{A43e} \\
\bar{D}_{1}C_{\mu \nu } &=&\bar{D}C_{\mu \nu }=\frac{1}{2}\partial _{\left[
\alpha \right. }C_{\left. \mu \right] \nu }dx^{\alpha },  \label{A43f} \\
\bar{D}_{1}\partial _{\left[ \alpha \right. }C_{\left. \mu \right] \nu }
&=&0,  \label{A43g} \\
\bar{D}_{1}\left( \gamma b\right) &=&0.  \label{A43h}
\end{eqnarray}
The nilpotency of $\bar{D}$ is equivalent to the nilpotency and the
anticommutation of its components
\begin{equation}
\bar{D}^{2}=0\Leftrightarrow \left( \bar{D}_{0}^{2}=0=\bar{D}_{1}^{2},\;\bar{%
D}_{0}\bar{D}_{1}+\bar{D}_{1}\bar{D}_{0}=0\right) .  \label{A44}
\end{equation}
We have to show that if $a$, with $\mathrm{pgh}\left( a\right) =2l>0$, $%
\mathrm{agh}\left( a\right) =k>0$ and $\deg \left( a\right) =p<\bar{D}$, is $%
\bar{D}$-closed, then it is $\bar{D}$-exact. Since $\bar{D}a=0$, we get that
$a$ is of the form (\ref{A34}), where there are precisely $\left( l+1\right)
$ terms in the sum from the right-hand side of (\ref{A34}), corresponding to
the $\left( l+1\right) $ elements $e^{J}$ of pure ghost number $2l$. We
reorganize $a$ like
\begin{equation}
a=\stackrel{(0)}{a}+\stackrel{(1)}{a}+\cdots +\stackrel{(l)}{a},  \label{A45}
\end{equation}
where the piece $\left( \stackrel{(i)}{a}\right) _{i=\overline{0,l}}$
contains $i$ antisymmetrized derivatives of the ghosts $\partial _{\left[
\alpha \right. }C_{\left. \mu \right] \nu }$ and $\left( l-i\right) $
undifferentiated ghosts $C_{\mu \nu }$ and call $\bar{D}$-degree the number
of factors of the type $\partial _{\left[ \alpha \right. }C_{\left. \mu
\right] \nu }$. It is clear from (\ref{A43a}--\ref{A43h}) that the action of
$\bar{D}_{0}$ on $\stackrel{(i)}{a}$ does not modify its $\bar{D}$-degree,
while the action of $\bar{D}_{1}$ on the same element increases its $\bar{D}$%
-degree by one unit. Using the expansion (\ref{A45}) and the decomposition (%
\ref{A42}), we get that the equation $\bar{D}a=0$ projected on the various
values of the $\bar{D}$-degree reads as
\begin{eqnarray}
\bar{D}_{0}\stackrel{(0)}{a} &=&0,  \label{A46a} \\
\bar{D}_{1}\stackrel{(0)}{a}+\bar{D}_{0}\stackrel{(1)}{a} &=&0,  \label{A46b}
\\
\bar{D}_{1}\stackrel{(i)}{a}+\bar{D}_{0}\stackrel{(i)}{a} &=&0,\;i=1,\cdots
,l-2,  \label{A46c} \\
\bar{D}_{1}\stackrel{(l-1)}{a}+\bar{D}_{0}\stackrel{(l)}{a} &=&0,
\label{A46d} \\
\bar{D}_{1}\stackrel{(l)}{a} &=&0.  \label{A46e}
\end{eqnarray}
The equation (\ref{A46e}) is satisfied due to the definitions (\ref{A43e})
and (\ref{A43g}), as $\stackrel{(l)}{a}$ contains only factors of the type $%
\partial _{\left[ \alpha \right. }C_{\left. \mu \right] \nu }$ and an
invariant polynomial. Denoting by $e^{J,i}$ the element of pure ghost number
$2l$ of a basis in the space of polynomials in $C_{\mu \nu }$ and $\partial
_{\left[ \alpha \right. }C_{\left. \mu \right] \nu }$ with the $\bar{D}$%
-degree equal to $i$, we have that
\begin{equation}
e^{J,i}\sim C_{\mu _{1}\nu _{1}}\cdots C_{\mu _{k-i}\nu _{k-i}}\partial
_{\left[ \alpha _{1}\right. }C_{\left. \beta _{1}\right] \gamma _{1}}\cdots
\partial _{\left[ \alpha _{i}\right. }C_{\left. \beta _{i}\right] \gamma
_{i}},  \label{A46f}
\end{equation}
such that
\begin{equation}
\stackrel{(i)}{a}=\sum_{J}\alpha _{J,i}e^{J,i},\;i=0,\cdots ,l,  \label{A46g}
\end{equation}
where each $\alpha _{J,i}$ is an invariant polynomial, with
\begin{equation}
\mathrm{agh}\left( \alpha _{J,i}\right) =k>0,\;\deg \left( \alpha
_{J,i}\right) =p<D,\;i=0,\cdots ,l.  \label{A46j}
\end{equation}
With this representation of the components of $a$ at hand and using the
definitions (\ref{A43a}--\ref{A43b}), the equation (\ref{A46a}) becomes $%
\left( d\alpha _{J,0}\right) e^{J,0}=0$, which is further equivalent with $%
d\alpha _{J,0}=0$ by the result (\ref{A36}). We are under the conditions of
Theorem \ref{hginv}, so
\begin{equation}
\alpha _{J,0}=d\beta _{J,0},\;\mathrm{agh}\left( \beta _{J,0}\right)
=k>0,\;\deg \left( \beta _{J,0}\right) =p-1,\;\mathrm{for\;all\;}J.
\label{A46k}
\end{equation}
Making the notation $\stackrel{(0)}{b}=\sum_{J}\beta _{J,0}e^{J,0}$, we find
that $a_{1}=a-\bar{D}\stackrel{(0)}{b}$ belongs to the same cohomological
class from $H\left( \bar{D}\right) $ like $a$. Moreover, $a_{1}$ starts from
the $\bar{D}$-degree equal to one
\begin{equation}
a_{1}=-\bar{D}_{1}\stackrel{(0)}{b}+\stackrel{(1)}{a}+\cdots +\stackrel{(l)}{%
a},  \label{A46l}
\end{equation}
such that the projection of the equation $\bar{D}a_{1}=0$ on the lowest
value of the $\bar{D}$-degree (one) reads as $\bar{D}_{0}\left( -\bar{D}_{1}%
\stackrel{(0)}{b}+\stackrel{(1)}{a}\right) =0$. By repeating the above
procedure, but with respect to $-\bar{D}_{1}\stackrel{(0)}{b}+\stackrel{(1)}{%
a}$, it follows that there exists an element $\stackrel{(1)}{b}%
=\sum_{J}\beta _{J,1}e^{J,1}$ such that $-\bar{D}_{1}\stackrel{(0)}{b}+%
\stackrel{(1)}{a}=\bar{D}_{0}\stackrel{(1)}{b}$, and hence $a_{2}=a_{1}-\bar{%
D}\stackrel{(1)}{b}=a-\bar{D}\left( \stackrel{(0)}{b}+\stackrel{(1)}{b}%
\right) $ lies in the same cohomological class from $H\left( \bar{D}\right) $
like $a$, but its decomposition begins with the $\bar{D}$-degree equal to
two
\begin{equation}
a_{2}=-\bar{D}_{1}\stackrel{(1)}{b}+\stackrel{(2)}{a}+\cdots +\stackrel{(l)}{%
a}.  \label{A46m}
\end{equation}
Reprising the same arguments, after $l$ steps we find that
\begin{equation}
a_{l}=a-\bar{D}\left( \stackrel{(0)}{b}+\stackrel{(1)}{b}+\cdots +\stackrel{%
(l-1)}{b}\right) =-\bar{D}_{1}\stackrel{(l-1)}{b}+\stackrel{(l)}{a},
\label{A47}
\end{equation}
whose $\bar{D}$-degree is equal to $l$, is equivalent with $a$ from the
point of view of the cohomology of $\bar{D}$. The equation $\bar{D}a_{l}=0$
reduces thus to
\begin{eqnarray}
\bar{D}_{0}\left( -\bar{D}_{1}\stackrel{(l-1)}{b}+\stackrel{(l)}{a}\right)
&=&0,  \label{A48a} \\
\bar{D}_{1}\left( -\bar{D}_{1}\stackrel{(l-1)}{b}+\stackrel{(l)}{a}\right)
&=&0.  \label{A48b}
\end{eqnarray}
The latter equation is automatically fulfilled due to (\ref{A43e}) and (\ref
{A43g}), while (\ref{A48a}) shows that there exists a $\stackrel{(l)}{b}$
such that
\begin{equation}
-\bar{D}_{1}\stackrel{(l-1)}{b}+\stackrel{(l)}{a}=\bar{D}_{0}\stackrel{(l)}{b%
}\equiv \bar{D}\stackrel{(l)}{b}.  \label{A49}
\end{equation}
Replacing (\ref{A49}) in (\ref{A47}), we have shown that
\begin{equation}
a=\bar{D}\left( \stackrel{(0)}{b}+\stackrel{(1)}{b}+\cdots +\stackrel{(l)}{b}%
\right) ,  \label{A49a}
\end{equation}
and hence $a$ is trivial in $H\left( \bar{D}\right) $. In conclusion, we can
state that $H\left( \bar{D}\right) $ is trivial at ghost number $g=2l-k$,
antighost number $k$ and form degree $p$, with $l,k>0$ and $p<D$, $%
H_{p}^{g,k}\left( \bar{D}\right) =0$, so by virtue of the relation (\ref
{A41a}) the proof of the theorem is now complete. $\blacksquare $

Theorem \ref{hdinhg} is one of the main tools needed for the computation of $%
H\left( s|d\right) $. In particular, it implies that there is no non-trivial
descent for $H\left( \gamma |d\right) $ in strictly positive antighost
number.

\begin{corollary}
\label{gaplusdb}If $a$ with
\begin{equation}
\mathrm{agh}\left( a\right) =k>0,\;\mathrm{gh}\left( a\right) =g\geq
-k,\;\deg \left( a\right) =p\leq D,  \label{A50a}
\end{equation}
satisfies the equation
\begin{equation}
\gamma a+db=0,  \label{A50}
\end{equation}
where
\begin{equation}
\mathrm{agh}\left( b\right) =k>0,\;\mathrm{gh}\left( b\right) =g+1>-k,\;\deg
\left( b\right) =p-1<D,  \label{A51a}
\end{equation}
then one can always redefine $a$%
\begin{equation}
a\rightarrow a^{\prime }=a+d\nu ,  \label{A51}
\end{equation}
so that
\begin{equation}
\gamma a^{\prime }=0.  \label{A52}
\end{equation}
\end{corollary}

\emph{Proof} We construct the descent associated with the equation (\ref{A50}%
). Acting with $\gamma $ on (\ref{A50}) and using the first and the third
relations in (\ref{A35a}), we find that
\begin{equation}
d\left( -\gamma b\right) =0,  \label{A53}
\end{equation}
such that the triviality of the cohomology of $d$ implies that
\begin{equation}
\gamma b+dc=0,  \label{A54}
\end{equation}
where
\begin{equation}
\mathrm{agh}\left( c\right) =k>0,\;\mathrm{gh}\left( c\right) =g+2,\;\deg
\left( c\right) =p-2.  \label{A55}
\end{equation}
Going on in the same way, we get the next equation from the descent
\begin{equation}
\gamma c+de=0,  \label{A56}
\end{equation}
with
\begin{equation}
\mathrm{agh}\left( e\right) =k>0,\;\mathrm{gh}\left( e\right) =g+3,\;\deg
\left( e\right) =p-3,  \label{A57}
\end{equation}
and so on. The descent stops after a finite number of steps with the last
equations
\begin{equation}
\gamma t+du=0,  \label{A57A}
\end{equation}
\begin{equation}
\gamma u+dv=0,  \label{A58}
\end{equation}
\begin{equation}
\gamma v=0,  \label{A59}
\end{equation}
either because $v$ is a zero-form or because we stopped at a higher
form-degree with a $\gamma $-closed term. It is essential to remark that
irrespective of the step at which the descent is cut, we have that
\begin{equation}
\mathrm{agh}\left( v\right) =k>0,\;\mathrm{gh}\left( v\right) =g^{\prime
}>-k,\;\deg \left( v\right) =p^{\prime }<D.  \label{A60}
\end{equation}
(The earliest step where the descent may terminate is $v=b$ and, according
to (\ref{A51a}), we have that $\deg \left( b\right) =p-1<D$ and $\mathrm{gh}%
\left( b\right) =g+1>-k$.)

The equations (\ref{A58}--\ref{A59}) together with the conditions (\ref{A60}%
) tell us that $v$ belongs to $H_{p^{\prime }}^{g^{\prime },k}\left(
d,H\left( \gamma \right) \right) $ for $k>0$, $p^{\prime }<D$ and $g^{\prime
}>-k$, so Theorem \ref{hdinhg} guarantees that $v$ is trivial in $%
H_{p^{\prime }}^{g^{\prime },k}\left( d,H\left( \gamma \right) \right) $%
\begin{equation}
v=d\nu ^{\prime }+\gamma \rho ^{\prime },\;\gamma \nu ^{\prime }=0,
\label{A61}
\end{equation}
which\footnote{%
Note that if the descent stops in form degree zero, $\deg \left( v\right) =0$%
, then the proof remains valid with the sole modification $\nu ^{\prime }=0$
in (\ref{A61}).} substituted in (\ref{A58}) allows us, due to the
anticommutation between $d$ and $\gamma $, to replace it with the equivalent
equation
\begin{equation}
\gamma u^{\prime }=0,  \label{A62}
\end{equation}
where
\begin{equation}
u^{\prime }=u-d\rho ^{\prime }.  \label{A63}
\end{equation}
In the meantime, (\ref{A63}) and the nilpotency of $d$ induces that $%
du^{\prime }=du$, such that the equation (\ref{A57A}) becomes
\begin{equation}
\gamma t+du^{\prime }=0.  \label{A64}
\end{equation}
Reprising the same argument in relation with (\ref{A62}) and the last
equation, we find that (\ref{A64}) can be replaced with
\begin{equation}
\gamma t^{\prime }=0,  \label{A65}
\end{equation}
where
\begin{equation}
t^{\prime }=t-d\rho ^{\prime \prime },  \label{A66}
\end{equation}
and $\rho ^{\prime \prime }$ comes from
\begin{equation}
u^{\prime }=d\nu ^{\prime \prime }+\gamma \rho ^{\prime \prime },\;\gamma
\nu ^{\prime \prime }=0.  \label{A67}
\end{equation}
Performing exactly the same operations for the remaining equations from the
descent, we finally infer that (\ref{A50}) is equivalent with
\begin{equation}
\gamma a^{\prime }=0,  \label{A68}
\end{equation}
where
\begin{equation}
a^{\prime }=a-d\rho ^{\prime \prime \prime },  \label{A69}
\end{equation}
and $\rho ^{\prime \prime \prime }$ appears in
\begin{equation}
b^{\prime }=d\nu ^{\prime \prime \prime }+\gamma \rho ^{\prime \prime \prime
},\;\gamma \nu ^{\prime \prime \prime }=0.  \label{A70}
\end{equation}
The corollary is now demonstrated once we perform the identification
\begin{equation}
\nu =-\rho ^{\prime \prime \prime },  \label{A71}
\end{equation}
between (\ref{A69}) and (\ref{A51}). Meanwhile, it is worth noticing that $%
b^{\prime }=b-dg$, with $\gamma g$ non-vanishing in general, so from (\ref
{A70}) we can also state that
\begin{equation}
b=\gamma \rho ^{\prime \prime \prime }+df,\;f=\nu ^{\prime \prime \prime }+g,
\label{A71a}
\end{equation}
with $\gamma f\neq 0$ in general. $\blacksquare $

\section{Some results on the (invariant) characteristic cohomology}

The second essential ingredient in the analysis of the local cohomology $%
H\left( s|d\right) $ is the local cohomology of the Koszul-Tate differential
in pure ghost number zero and in strictly positive antighost numbers, $%
H\left( \delta |d\right) $, also known as the characteristic cohomology. We
recall that the local cohomology $H\left( \delta |d\right) $ is completely
trivial at both strictly positive antighost \textit{and} pure ghost numbers
(for instance, see~\cite{gen1}, Theorem 5.4 and~\cite{commun1}). An element $%
\alpha $ with the properties
\begin{equation}
\mathrm{agh}\left( \alpha \right) >0,\;\mathrm{pgh}\left( \alpha \right) =0,
\label{propdelta}
\end{equation}
is said to belong to $H\left( \delta |d\right) $ if and only if it is $%
\delta $ closed modulo $d$
\begin{equation}
\delta \alpha =dj,\;\mathrm{pgh}\left( j\right) =0.  \label{r80}
\end{equation}
If $\alpha \in H\left( \delta |d\right) $ is a $\delta $-boundary modulo $d$%
\begin{equation}
\alpha =\delta b+dc,\;\mathrm{pgh}\left( \alpha \right) =\mathrm{pgh}\left(
\beta \right) =\mathrm{pgh}\left( b\right) =0,\;\mathrm{agh}\left( \alpha
\right) =\mathrm{agh}\left( b\right) >0,  \label{r80a}
\end{equation}
we will call it trivial in $H\left( \delta |d\right) $. The solution to the
equation (\ref{r80}) is thus unique up to trivial objects, $\alpha
\rightarrow \alpha +\delta b+dc$. The local cohomology $H\left( \delta
|d\right) $ inherits a natural grading in terms of the antighost number,
such that from now on we will denote by $H_{k}\left( \delta |d\right) $ the
local cohomology of $\delta $ in antighost number $k$. As we have discussed
in Section \ref{2}, the free model under study is a normal gauge theory of
Cauchy order equal to three. Using the general results from~\cite{gen1}
(also see~\cite{lingr} and~\cite{multi,gen2}), one can state that the local
cohomology of the Koszul-Tate differential at pure ghost number zero is
trivial in antighost numbers strictly greater than its Cauchy order
\begin{equation}
H_{k}\left( \delta |d\right) =0,\;k>3.  \label{r81}
\end{equation}

The final tool needed for the calculation of $H\left( s|d\right) $ is the
local cohomology of the Koszul-Tate differential in the space of invariant
polynomials, $H^{\mathrm{inv}}\left( \delta |d\right) $, also called the
invariant characteristic cohomology. It is defined via an equation similar
to (\ref{r80}), but with $\alpha $ and $j$ replaced by invariant
polynomials. Along the same line, the notion of trivial element from $H^{%
\mathrm{inv}}\left( \delta |d\right) $ is revealed by (\ref{r80a}) up to the
precaution that both $b$ and $c$ must be invariant polynomials. It appears
the natural question if the result (\ref{r81}) is still valid in the space
of invariant polynomials. The answer is affirmative
\begin{equation}
H_{k}^{\mathrm{inv}}\left( \delta |d\right) =0,\;k>3  \label{r81c}
\end{equation}
and is proved below, in Theorem \ref{hinvdelta}. Actually, we prove that if $%
\alpha _{k}$ is trivial in $H_{k}\left( \delta |d\right) $, then it can be
taken to be trivial also in $H_{k}^{\mathrm{inv}}\left( \delta |d\right) $.
We consider only the case $k\geq 3$ since our main scope is to argue the
triviality of $H^{\mathrm{inv}}\left( \delta |d\right) $ in antighost number
strictly greater than three. First, we prove the following lemma.

\begin{lemma}
\label{trivinv}Let $\alpha $ be a $\delta $-exact invariant polynomial
\begin{equation}
\alpha =\delta \beta .  \label{A72}
\end{equation}
Then, $\beta $ can also be taken to be an invariant polynomial.
\end{lemma}

\emph{Proof} Let $v$ be a function of $\left[ \chi ^{*\Delta }\right] $ and $%
\left[ t_{\mu \nu |\alpha \beta }\right] $. The dependence of $v$ on $\left[
t_{\mu \nu |\alpha \beta }\right] $ can be reorganized as a dependence on
the curvature and its derivatives, $\left[ F\right] $, and on $\tilde{t}%
_{\mu \nu |\alpha \beta }=\left\{ t_{\mu \nu |\alpha \beta },\partial t_{\mu
\nu |\alpha \beta },\cdots \right\} $, where $\tilde{t}_{\mu \nu |\alpha
\beta }$ are not $\gamma $-invariant. If $v$ is $\gamma $-invariant, then it
does not involve $\tilde{t}_{\mu \nu |\alpha \beta }$, i.e., $v=\left.
v\right| _{\tilde{t}_{\mu \nu |\alpha \beta }=0}$, so we have by hypothesis
that
\begin{equation}
\alpha =\left. \alpha \right| _{\tilde{t}_{\mu \nu |\alpha \beta }=0}.
\label{A73}
\end{equation}
On the other hand, $\beta $ depends in general on $\left[ \chi ^{*\Delta
}\right] $, $\left[ F\right] $ and $\tilde{t}_{\mu \nu |\alpha \beta }$.
Making $\tilde{t}_{\mu \nu |\alpha \beta }=0$ in (\ref{A72}), using (\ref
{A73}) and taking into account the fact that $\delta $ commutes with the
operation of setting $\tilde{t}_{\mu \nu |\alpha \beta }$ equal to zero, we
find that
\begin{equation}
\alpha =\delta \left( \left. \beta \right| _{\tilde{t}_{\mu \nu |\alpha
\beta }=0}\right) ,  \label{A74}
\end{equation}
with $\left. \beta \right| _{\tilde{t}_{\mu \nu |\alpha \beta }=0}$
invariant. This proves the lemma. $\blacksquare $

Now, we have the necessary tools for proving the next theorem.

\begin{theorem}
\label{hinvdelta}Let $\alpha _{k}^{p}$ be an invariant polynomial with $\deg
\left( \alpha _{k}^{p}\right) =p$ and $\mathrm{agh}\left( \alpha
_{k}^{p}\right) =k$, which is $\delta $-exact modulo $d$%
\begin{equation}
\alpha _{k}^{p}=\delta \lambda _{k+1}^{p}+d\lambda _{k}^{p-1},\;k\geq 3.
\label{A75}
\end{equation}
Then, we can choose $\lambda _{k+1}^{p}$ and $\lambda _{k}^{p-1}$ to be
invariant polynomials.
\end{theorem}

\emph{Proof} Initially, by successively acting with $d$ and $\delta $ on (%
\ref{A75}) we obtain a tower of equations of the same type. Indeed, acting
with $d$ on (\ref{A75}) we find that $d\alpha _{k}^{p}=-\delta \left(
d\lambda _{k+1}^{p}\right) $. On the other hand, as $d\alpha _{k}^{p}$ is
invariant, by means of Lemma \ref{trivinv} we obtain that $d\alpha
_{k}^{p}=-\delta \alpha _{k+1}^{p+1}$, with $\alpha _{k+1}^{p+1}$ invariant.
From the last two relations we deduce that $\delta \left( \alpha
_{k+1}^{p+1}-d\lambda _{k+1}^{p}\right) =0$. As $\delta $ is acyclic at
strictly positive antighost numbers, the last relation implies that
\begin{equation}
\alpha _{k+1}^{p+1}=\delta \lambda _{k+2}^{p+1}+d\lambda _{k+1}^{p}.
\label{A76}
\end{equation}
Starting now with (\ref{A76}) and reprising the same operations like those
performed between the formulas (\ref{A75}) and (\ref{A76}), we obtain a
descent that stops in form degree $D$ with the equation $\alpha
_{k+D-p}^{D}=\delta \lambda _{k+D-p+1}^{D}+d\lambda _{k+D-p}^{D-1}$. Now, we
act with $\delta $ on (\ref{A75}) and deduce that $\delta \alpha
_{k}^{p}=-d\delta \lambda _{k}^{p-1}$. As $\delta \alpha _{k}^{p}$ is
invariant, in the case $k>1$, due to the Theorem \ref{hginv}, we obtain that
$\delta \alpha _{k}^{p}=-d\alpha _{k-1}^{p-1}$, where $\alpha _{k-1}^{p-1}$
is invariant. Using the last two relations we get that $d\left( \alpha
_{k-1}^{p-1}-\delta \lambda _{k}^{p-1}\right) =0$, such that it follows that
\begin{equation}
\alpha _{k-1}^{p-1}=\delta \lambda _{k}^{p-1}+d\lambda _{k-1}^{p-2}.
\label{A77}
\end{equation}
If $k=3$ in (\ref{A75}), we cannot go down since by assumption $k\geq 3$,
and so the bottom of the tower is (\ref{A75}) for $k=3$. Starting from (\ref
{A77}) and reprising the same procedure we reach a descent that ends at
either form degree zero or antighost number three, hence the last equation
respectively takes the form
\begin{equation}
\alpha _{k-p}^{0}=\delta \lambda _{k-p+1}^{0},  \label{A78}
\end{equation}
for $k-p\geq 3$ or
\begin{equation}
\alpha _{3}^{p-k+3}=\delta \lambda _{4}^{p-k+3}+d\lambda _{3}^{p-k+2},
\label{A79}
\end{equation}
for $k-p<3$. In consequence, the procedure described in the above leads to
the chain
\[
\alpha _{k+D-p}^{D}=\delta \lambda _{k+D-p+1}^{D}+d\lambda _{k+D-p}^{D-1},
\]
\[
\vdots
\]
\[
\alpha _{k+1}^{p+1}=\delta \lambda _{k+2}^{p+1}+d\lambda _{k+1}^{p},
\]
\[
\alpha _{k}^{p}=\delta \lambda _{k+1}^{p}+d\lambda _{k}^{p-1},
\]
\[
\alpha _{k-1}^{p-1}=\delta \lambda _{k}^{p-1}+d\lambda _{k-1}^{p-2},
\]
\[
\vdots
\]
\begin{equation}
\alpha _{k-p}^{0}=\delta \lambda _{k-p+1}^{0}\;\mathrm{or}\;\alpha
_{3}^{p-k+3}=\delta \lambda _{4}^{p-k+3}+d\lambda _{3}^{p-k+2}.  \label{A80}
\end{equation}
All the $\alpha $'s in the descent (\ref{A80}) are invariant.

Now, we show that if one of the $\lambda $'s in (\ref{A80}) is invariant,
then all the other $\lambda $'s can be taken to be also invariant. Indeed,
let $\lambda _{B}^{A-1}$ be invariant. It is involved in two of the
equations from (\ref{A80}), namely
\begin{equation}
\alpha _{B}^{A}=\delta \lambda _{B+1}^{A}+d\lambda _{B}^{A-1},  \label{A81}
\end{equation}
\begin{equation}
\alpha _{B-1}^{A-1}=\delta \lambda _{B}^{A-1}+d\lambda _{B-1}^{A-2}.
\label{A82}
\end{equation}
From (\ref{A81}) it results that $\alpha _{B}^{A}-d\lambda _{B}^{A-1}$ is
invariant. Then, in agreement with Lemma \ref{trivinv} the object $\lambda
_{B+1}^{A}$ can be chosen to be invariant. Using (\ref{A82}), we have that $%
\alpha _{B-1}^{A-1}-\delta \lambda _{B}^{A-1}$ is invariant, such that
Theorem \ref{hginv} ensures that $\lambda _{B-1}^{A-2}$ is also invariant.
On the other hand, $\lambda _{B+1}^{A}$ and $\lambda _{B-1}^{A-2}$ are
involved in other two sets of equations from the descent. (For instance, the
former element appears in the equations $\alpha _{B+1}^{A+1}=\delta \lambda
_{B+2}^{A+1}+d\lambda _{B+1}^{A}$ and $\alpha _{B+2}^{A+2}=\delta \lambda
_{B+3}^{A+2}+d\lambda _{B+2}^{A+1}$.) Going on in the same fashion, we find
that all the $\lambda $'s are invariant. In the case where $\lambda
_{B}^{A-1}$ appears at the top or at the bottom of the descent, we act in a
similar way, but only with respect to a single equation. The above
considerations emphasize that it is enough to verify the theorem in form
degree $D$ and for all the values $k\geq 3$ of the antighost number.

If $k\geq D+3$ (and hence $k-p\geq 3$), the last equation from the descent (%
\ref{A80}) for $p=D$ reads as
\begin{equation}
\alpha _{k-D}^{0}=\delta \lambda _{k-D+1}^{0}.  \label{A83}
\end{equation}
Using Lemma \ref{trivinv}, it results that $\lambda _{k-D+1}^{0}$ can be
taken to be invariant, such that the above arguments lead to the conclusion
that all the $\lambda $'s from the descent can be chosen invariant. As a
consequence, in the first equation from the descent in this situation,
namely, $\alpha _{k}^{D}=\delta \lambda _{k+1}^{D}+d\lambda _{k}^{D-1}$, we
have that both $\lambda _{k+1}^{D}$ and $\lambda _{k}^{D-1}$ are invariant.
Therefore, the theorem is true in form degree $D$ and in all antighost
numbers $k\geq D+3$, so it remains to be proved that it holds in form degree
$D$ and in all antighost numbers $3\leq k<D+3$. This is done below.

In the sequel we consider the case $p=D$ and $3\leq k<D+3$. The top equation
from (\ref{A80}), written in dual notations, takes the form
\begin{equation}
\alpha _{k}=\delta \lambda _{k+1}+\partial _{\mu }\lambda _{k}^{\mu
},\;3\leq k<D+3.  \label{A84}
\end{equation}
On the other hand, we can express $\alpha _{k}$ in terms of its E.L.
derivatives by means of the homotopy formula
\begin{eqnarray}
\alpha _{k} &=&\int\nolimits_{0}^{1}d\tau \left( \frac{\delta ^{R}\alpha _{k}%
}{\delta C_{\mu \nu }^{*}}\left( \tau \right) C_{\mu \nu }^{*}+\frac{\delta
^{R}\alpha _{k}}{\delta \eta _{\mu \nu |\alpha }^{*}}\left( \tau \right)
\eta _{\mu \nu |\alpha }^{*}\right.  \nonumber \\
&&\left. +\frac{\delta ^{R}\alpha _{k}}{\delta t_{\mu \nu |\alpha \beta }^{*}%
}\left( \tau \right) t_{\mu \nu |\alpha \beta }^{*}+\frac{\delta ^{R}\alpha
_{k}}{\delta t_{\mu \nu |\alpha \beta }}\left( \tau \right) t_{\mu \nu
|\alpha \beta }\right) +\partial _{\mu }j_{k}^{\mu },  \label{A85}
\end{eqnarray}
where $\frac{\delta ^{R}\alpha _{k}}{\delta C_{\mu \nu }^{*}}\left( \tau
\right) =\frac{\delta ^{R}\alpha _{k}}{\delta C_{\mu \nu }^{*}}\left( \tau
\left[ t_{\mu \nu |\alpha \beta }\right] ,\tau \left[ \chi ^{*\Delta
}\right] \right) $ and similarly for the other terms. Denoting the E.L.
derivatives of $\lambda _{k+1}$ by
\begin{equation}
\frac{\delta ^{R}\lambda _{k+1}}{\delta C_{\mu \nu }^{*}}=G_{k-2}^{\mu \nu
},\;\frac{\delta ^{R}\lambda _{k+1}}{\delta \eta _{\mu \nu |\alpha }^{*}}%
=M_{k-1}^{\mu \nu |\alpha },  \label{A86}
\end{equation}
\begin{equation}
\frac{\delta ^{R}\lambda _{k+1}}{\delta t_{\mu \nu |\alpha \beta }^{*}}%
=N_{k}^{\mu \nu |\alpha \beta },\;\frac{\delta ^{R}\lambda _{k+1}}{\delta
t_{\mu \nu |\alpha \beta }}=L_{k+1}^{\mu \nu |\alpha \beta },  \label{A87}
\end{equation}
and using (\ref{A84}) we find after some computation that the E.L.
derivatives of $\alpha _{k}$ are given by
\begin{equation}
\frac{\delta ^{R}\alpha _{k}}{\delta C_{\mu \nu }^{*}}=-\delta G_{k-2}^{\mu
\nu },\;\frac{\delta ^{R}\alpha _{k}}{\delta \eta _{\mu \nu |\alpha }^{*}}%
=\delta M_{k-1}^{\mu \nu |\alpha }+\partial ^{\left[ \mu \right.
}G_{k-2}^{\left. \nu \right] \alpha }-2\partial ^{\alpha }G_{k-2}^{\mu \nu },
\label{A88}
\end{equation}
\begin{equation}
\frac{\delta ^{R}\alpha _{k}}{\delta t_{\mu \nu |\alpha \beta }^{*}}=-\delta
N_{k}^{\mu \nu |\alpha \beta }+\partial ^{\alpha }M_{k-1}^{\mu \nu |\beta
}-\partial ^{\beta }M_{k-1}^{\mu \nu |\alpha }+\partial ^{\mu
}M_{k-1}^{\alpha \beta |\nu }-\partial ^{\nu }M_{k-1}^{\alpha \beta |\mu },
\label{A89}
\end{equation}
\begin{equation}
\frac{\delta ^{R}\alpha _{k}}{\delta t_{\mu \nu |\alpha \beta }}=\delta
L_{k+1}^{\mu \nu |\alpha \beta }+\frac{1}{8}\partial _{\rho }\partial
_{\gamma }N_{k}^{\mu \nu \rho |\alpha \beta \gamma },  \label{A90}
\end{equation}
where $N_{k}^{\mu \nu \rho |\alpha \beta \gamma }$ has the same mixed
symmetry like the curvature tensor $F^{\mu \nu \rho |\alpha \beta \gamma }$%
\begin{eqnarray}
N_{k}^{\mu \nu \rho |\alpha \beta \gamma } &=&\sigma ^{\gamma \left[ \rho
\right. }N_{k}^{\left. \mu \nu \right] |\alpha \beta }+\sigma ^{\alpha
\left[ \rho \right. }N_{k}^{\left. \mu \nu \right] |\beta \gamma }+\sigma
^{\beta \left[ \rho \right. }N_{k}^{\left. \mu \nu \right] |\gamma \alpha
}+\sigma ^{\rho \left[ \gamma \right. }N_{k}^{\left. \alpha \beta \right]
|\mu \nu }  \nonumber \\
&&+\sigma ^{\mu \left[ \gamma \right. }N_{k}^{\left. \alpha \beta \right]
|\nu \rho }+\sigma ^{\nu \left[ \gamma \right. }N_{k}^{\left. \alpha \beta
\right] |\rho \mu }-2\left( \sigma ^{\gamma \left[ \rho \right. }\sigma
^{\left. \mu \right] \alpha }N_{k}^{\beta \nu }\right.  \nonumber \\
&&+\sigma ^{\gamma \left[ \mu \right. }\sigma ^{\left. \nu \right] \alpha
}N_{k}^{\beta \rho }+\sigma ^{\gamma \left[ \nu \right. }\sigma ^{\left.
\rho \right] \alpha }N_{k}^{\beta \mu }+\sigma ^{\alpha \left[ \rho \right.
}\sigma ^{\left. \mu \right] \beta }N_{k}^{\gamma \nu }  \nonumber \\
&&+\sigma ^{\alpha \left[ \mu \right. }\sigma ^{\left. \nu \right] \beta
}N_{k}^{\gamma \rho }+\sigma ^{\alpha \left[ \nu \right. }\sigma ^{\left.
\rho \right] \beta }N_{k}^{\gamma \mu }+\sigma ^{\beta \left[ \rho \right.
}\sigma ^{\left. \mu \right] \gamma }N_{k}^{\alpha \nu }  \nonumber \\
&&\left. +\sigma ^{\beta \left[ \mu \right. }\sigma ^{\left. \nu \right]
\gamma }N_{k}^{\alpha \rho }+\sigma ^{\beta \left[ \nu \right. }\sigma
^{\left. \rho \right] \gamma }N_{k}^{\alpha \mu }\right) +\left( \sigma
^{\gamma \left[ \rho \right. }\sigma ^{\left. \mu \right] \alpha }\sigma
^{\beta \nu }\right.  \nonumber \\
&&\left. +\sigma ^{\gamma \left[ \mu \right. }\sigma ^{\left. \nu \right]
\alpha }\sigma ^{\beta \rho }+\sigma ^{\gamma \left[ \nu \right. }\sigma
^{\left. \rho \right] \alpha }\sigma ^{\beta \mu }\right) N_{k},
\label{A90a}
\end{eqnarray}
and we employed the notations $N_{k}^{\nu \beta }=\sigma _{\mu \alpha
}N_{k}^{\mu \nu |\alpha \beta }$ and $N_{k}=\sigma _{\nu \beta }N_{k}^{\nu
\beta }$. As the E.L. derivatives of an invariant quantity are also
invariant, the former equation in (\ref{A88}) together with Lemma \ref
{trivinv} (as $k-2>0$) lead to
\begin{equation}
\frac{\delta ^{R}\alpha _{k}}{\delta C_{\mu \nu }^{*}}=-\delta \bar{G}%
_{k-2}^{\mu \nu },  \label{A91}
\end{equation}
with $\bar{G}_{k-2}^{\mu \nu }$ invariant. Following a similar reasoning, we
find that
\begin{equation}
\frac{\delta ^{R}\alpha _{k}}{\delta \eta _{\mu \nu |\alpha }^{*}}=\delta
\bar{M}_{k-1}^{\mu \nu |\alpha }+\left( \partial ^{\left[ \mu \right. }\bar{G%
}_{k-2}^{\left. \nu \right] \alpha }-2\partial ^{\alpha }\bar{G}_{k-2}^{\mu
\nu }\right) ,  \label{A92}
\end{equation}
\begin{equation}
\frac{\delta ^{R}\alpha _{k}}{\delta t_{\mu \nu |\alpha \beta }^{*}}=-\delta
\bar{N}_{k}^{\mu \nu |\alpha \beta }+\partial ^{\alpha }\bar{M}_{k-1}^{\mu
\nu |\beta }-\partial ^{\beta }\bar{M}_{k-1}^{\mu \nu |\alpha }+\partial
^{\mu }\bar{M}_{k-1}^{\alpha \beta |\nu }-\partial ^{\nu }\bar{M}%
_{k-1}^{\alpha \beta |\mu },  \label{A93}
\end{equation}
\begin{equation}
\frac{\delta ^{R}\alpha _{k}}{\delta t_{\mu \nu |\alpha \beta }}=\delta \bar{%
L}_{k+1}^{\mu \nu |\alpha \beta }+\frac{1}{8}\partial _{\rho }\partial
_{\gamma }\bar{N}_{k}^{\mu \nu \rho |\alpha \beta \gamma },  \label{A94}
\end{equation}
where all the bar quantities are invariant. Since $\alpha _{k}$ is
invariant, it depends on $t_{\mu \nu |\alpha \beta }$ only through the
curvature and its derivatives, such that
\begin{equation}
\frac{\delta ^{R}\alpha _{k}}{\delta t_{\mu \nu |\alpha \beta }}=\partial
_{\rho }\partial _{\gamma }\Delta _{k}^{\mu \nu \rho |\alpha \beta \gamma },
\label{A95}
\end{equation}
where $\Delta _{k}^{\mu \nu \rho |\alpha \beta \gamma }$ has the mixed
symmetry of the curvature tensor. The part from (\ref{A94}) involving $\bar{N%
}$ has a form similar to that of the right-hand side of (\ref{A95}). Then, $%
\delta \bar{L}_{k+1}^{\mu \nu |\alpha \beta }$ must be expressed in the same
manner, i.e.,
\begin{equation}
\delta \bar{L}_{k+1}^{\mu \nu |\alpha \beta }=\partial _{\rho }\partial
_{\gamma }\Omega _{k}^{\mu \nu \rho |\alpha \beta \gamma },  \label{A96}
\end{equation}
for some $\Omega $ with the mixed symmetry of the curvature. The equation (%
\ref{A96}) shows that for some given $\alpha $ and $\beta $, the object $%
\bar{L}_{k+1}^{\mu \nu |\alpha \beta }$ belongs to $H_{k+1}^{D-2}\left(
\delta |d\right) $. As $H_{k+1}^{D-2}\left( \delta |d\right) \simeq
H_{k+2}^{D-1}\left( \delta |d\right) \simeq H_{k+3}^{D}\left( \delta
|d\right) $ (see \cite{gen1}, Theorem 8.1) and $H_{k+3}^{D}\left( \delta
|d\right) \simeq 0$, the equation (\ref{A96}) implies that
\begin{equation}
\bar{L}_{k+1}^{\mu \nu |\alpha \beta }=\delta R_{k+2}^{\mu \nu \alpha \beta
}+\partial _{\rho }U_{k+1}^{\rho \mu \nu \alpha \beta },  \label{A97}
\end{equation}
where $R_{k+2}^{\mu \nu \alpha \beta }$ is separately antisymmetric in $%
\left\{ \mu ,\nu \right\} $ and $\left\{ \alpha ,\beta \right\} $, and $%
U_{k+1}^{\rho \mu \nu \alpha \beta }$ is antisymmetric in $\left\{ \rho ,\mu
,\nu \right\} $, as well as in $\left\{ \alpha ,\beta \right\} $.

Now, we prove the theorem in the case $3\leq k<D+3$ by induction. This is,
we assume that the theorem is valid in antighost number $\left( k+3\right) $
and in form degree $D$, and show that it holds in antighost number $k$ and
in form degree $D$. In agreement with the induction hypothesis, $%
R_{k+2}^{\mu \nu \alpha \beta }$ and $U_{k+1}^{\rho \mu \nu \alpha \beta }$
can be assumed to be invariant. On the other hand, $\bar{L}_{k+1}^{\mu \nu
|\alpha \beta }$ must verify the mixed symmetry of the tensor field $t^{\mu
\nu |\alpha \beta }$ with respect to the given values $\alpha $ and $\beta $%
, i.e., $\bar{L}_{k+1}^{\mu \left[ \nu |\alpha \beta \right] }=0$, which
further implies that
\begin{equation}
\delta R_{k+2}^{\mu \left[ \nu \alpha \beta \right] }+\partial _{\rho
}U_{k+1}^{\rho \mu \left[ \nu \alpha \beta \right] }=0.  \label{A98}
\end{equation}
Acting with $\delta $ on (\ref{A98}), we obtain $\partial _{\rho }\left(
\delta U_{k+1}^{\rho \mu \left[ \nu \alpha \beta \right] }\right) =0$, such
that
\begin{equation}
\delta U_{k+1}^{\rho \mu \left[ \nu \alpha \beta \right] }=\partial _{\gamma
}V_{k}^{\gamma \rho \mu ||\nu \alpha \beta },  \label{A99}
\end{equation}
where $V_{k}^{\gamma \rho \mu ||\nu \alpha \beta }$ is separately
antisymmetric in $\left\{ \gamma ,\rho ,\mu \right\} $ and $\left\{ \nu
\alpha \beta \right\} $ (the double bar $||$ signifies that in general $%
V_{k+1}^{\gamma \rho \mu ||\nu \alpha \beta }$ neither satisfies the
identity $V_{k+1}^{\left[ \gamma \rho \mu ||\nu \right] \alpha \beta }\equiv
0$ nor is symmetric under the permutation $\gamma \longleftrightarrow \nu $,
$\rho \longleftrightarrow \alpha $, $\mu \longleftrightarrow \beta $). The
equation (\ref{A99}) shows that for some fixed $\nu $, $\alpha $ and $\beta $%
, $U_{k+1}^{\rho \mu \left[ \nu \alpha \beta \right] }$ pertains to $%
H_{k+1}^{D-2}\left( \delta |d\right) $, that is finally found isomorphic to $%
H_{k+3}^{D}\left( \delta |d\right) \simeq 0$, so $U_{k+1}^{\rho \mu \left[
\nu \alpha \beta \right] }$ is trivial
\begin{equation}
U_{k+1}^{\rho \mu \left[ \nu \alpha \beta \right] }=\delta W_{k+2}^{\rho \mu
\nu \alpha \beta }+\partial _{\gamma }S_{k+1}^{\gamma \rho \mu ||\nu \alpha
\beta },  \label{A100}
\end{equation}
with $W_{k+2}^{\rho \mu \nu \alpha \beta }$ antisymmetric in both $\left\{
\rho ,\mu \right\} $ and $\left\{ \nu ,\alpha ,\beta \right\} $ and $%
S_{k+1}^{\gamma \rho \mu ||\nu \alpha \beta }$ separately antisymmetric in $%
\left\{ \gamma ,\rho ,\mu \right\} $ and $\left\{ \nu ,\alpha ,\beta
\right\} $. Using again the induction hypothesis, we can assume that $%
W_{k+2}^{\rho \mu \nu \alpha \beta }$ and $S_{k+1}^{\gamma \rho \mu ||\nu
\alpha \beta }$ are invariant. In order to reconstruct $\alpha _{k}$ through
the homotopy formula (\ref{A85}), we need to compute $\delta \bar{L}%
_{k+1}^{\mu \nu |\alpha \beta }$ by means of formula (\ref{A97}), so
eventually we need to calculate $\partial _{\rho }U_{k+1}^{\rho \mu \nu
\alpha \beta }$. In this respect we use the equation (\ref{A100}) and the
identity (that holds only for a tensor that is separately antisymmetric in
its first three and respectively in its last two indices)
\begin{eqnarray}
U_{k+1}^{\rho \mu \nu \alpha \beta } &=& \frac{1}{6}\left( 2\left(
U_{k+1}^{\rho \mu \left[ \nu \alpha \beta \right] }+U_{k+1}^{\mu \nu \left[
\rho \alpha \beta \right] }+U_{k+1}^{\nu \rho \left[ \mu \alpha \beta
\right] }+U_{k+1}^{\alpha \beta \left[ \rho \mu \nu \right] }\right) \right.
\nonumber \\
&&+U_{k+1}^{\rho \beta \left[ \alpha \mu \nu \right] }-U_{k+1}^{\rho \alpha
\left[ \beta \mu \nu \right] }+U_{k+1}^{\mu \beta \left[ \alpha \nu \rho
\right] }  \nonumber \\
&&\left. -U_{k+1}^{\mu \alpha \left[ \beta \nu \rho \right] }+U_{k+1}^{\nu
\beta \left[ \alpha \rho \mu \right] }-U_{k+1}^{\nu \alpha \left[ \beta \rho
\mu \right] }\right) ,  \label{A101}
\end{eqnarray}
and obtain that
\begin{equation}
\partial _{\rho }U_{k+1}^{\rho \mu \nu \alpha \beta }=\delta \tilde{W}%
_{k+2}^{\mu \nu \alpha \beta }+\partial _{\rho }\partial _{\gamma }\left(
G_{k+1}^{\mu \nu \rho \alpha \beta \gamma }+E_{k+1}^{\mu \nu \rho \alpha
\beta \gamma }\right) ,  \label{A102}
\end{equation}
where
\begin{equation}
G_{k+1}^{\mu \nu \rho \alpha \beta \gamma }=\frac{1}{4}\left( S_{k+1}^{\mu
\nu \rho ||\alpha \beta \gamma }+S_{k+1}^{\alpha \beta \gamma ||\mu \nu \rho
}\right) ,  \label{A102a}
\end{equation}
\begin{eqnarray}
E_{k+1}^{\mu \nu \rho \alpha \beta \gamma } &=&\frac{1}{12}\left(
S_{k+1}^{\beta \gamma \left[ \mu ||\nu \rho \right] \alpha }+S_{k+1}^{\gamma
\alpha \left[ \mu ||\nu \rho \right] \beta }+S_{k+1}^{\alpha \beta \left[
\mu ||\nu \rho \right] \gamma }\right.  \nonumber \\
&&\left. +S_{k+1}^{\nu \rho \left[ \alpha ||\beta \gamma \right] \mu
}+S_{k+1}^{\rho \mu \left[ \alpha ||\beta \gamma \right] \nu }+S_{k+1}^{\mu
\nu \left[ \alpha ||\beta \gamma \right] \rho }\right) .  \label{A103}
\end{eqnarray}
Obviously, $G_{k+1}^{\mu \nu \rho \alpha \beta \gamma }$ and $E_{k+1}^{\mu
\nu \rho \alpha \beta \gamma }$ are invariant as $S_{k+1}^{\mu \nu \rho
||\alpha \beta \gamma }$ is invariant. By direct handling it can be shown
that $G_{k+1}^{\mu \nu \rho \alpha \beta \gamma }=G_{k+1}^{\mu \nu \rho
|\alpha \beta \gamma }$ and $E_{k+1}^{\mu \nu \rho \alpha \beta \gamma
}=E_{k+1}^{\mu \nu \rho |\alpha \beta \gamma }$ in the sense that they
indeed display the mixed symmetry of the curvature tensor (they are
separately antisymmetric in the indices $\left\{ \mu ,\nu ,\rho \right\} $
and $\left\{ \alpha ,\beta ,\gamma \right\} $ and also symmetric under the
interchange $\left\{ \mu ,\nu ,\rho \right\} \longleftrightarrow \left\{
\alpha ,\beta ,\gamma \right\} $, although they do not verify in general the
Bianchi I identity $G_{k+1}^{\left[ \mu \nu \rho |\alpha \right] \beta
\gamma }\equiv 0$ or $E_{k+1}^{\left[ \mu \nu \rho |\alpha \right] \beta
\gamma }\equiv 0$). Denoting by $\tilde{S}_{k+1}^{\mu \nu \rho |\alpha \beta
\gamma }$ the term $G_{k+1}^{\mu \nu \rho |\alpha \beta \gamma
}+E_{k+1}^{\mu \nu \rho |\alpha \beta \gamma }$, it is then obvious that it
is invariant and possesses the mixed symmetry of the curvature tensor (in
the sense specified in the above). Using the above notation and inserting (%
\ref{A102}) in (\ref{A97}), it results that
\begin{equation}
\bar{L}_{k+1}^{\mu \nu |\alpha \beta }=\delta \tilde{R}_{k+2}^{\mu \nu
\alpha \beta }+\partial _{\rho }\partial _{\gamma }\tilde{S}_{k+1}^{\mu \nu
\rho |\alpha \beta \gamma }.  \label{A104}
\end{equation}
With the help of (\ref{A91}--\ref{A94}) and (\ref{A104}), the formula (\ref
{A85}) becomes
\begin{eqnarray}
\alpha _{k} &=&\delta \left[ \int\nolimits_{0}^{1}d\tau \left( \bar{G}%
_{k-2}^{\mu \nu }C_{\mu \nu }^{*}+\bar{M}_{k-1}^{\mu \nu |\alpha }\eta _{\mu
\nu |\alpha }^{*}+\bar{N}_{k}^{\mu \nu |\alpha \beta }t_{\mu \nu |\alpha
\beta }^{*}\right. \right.  \nonumber \\
&&\left. \left. +\left( \partial _{\rho }\partial _{\gamma }\tilde{S}
_{k+1}^{\mu \nu \rho |\alpha \beta \gamma }\right) t_{\mu \nu |\alpha \beta
}\right) \right] +\partial _{\mu }\sigma _{k}^{\mu }.  \label{A105}
\end{eqnarray}
The last term in the argument of $\delta $ can be written in the form
\begin{equation}
\left( \partial _{\rho }\partial _{\gamma }\tilde{S}_{k+1}^{\mu \nu \rho
|\alpha \beta \gamma }\right) t_{\mu \nu |\alpha \beta }=\frac{1}{9}\tilde{S}%
_{k+1}^{\mu \nu \rho |\alpha \beta \gamma }F_{\mu \nu \rho |\alpha \beta
\gamma }+\partial _{\mu }\phi _{k+1}^{\mu },  \label{A106}
\end{equation}
so finally we arrive at
\begin{eqnarray}
\alpha _{k} &=&\delta \left[ \int\nolimits_{0}^{1}d\tau \left( \bar{G}%
_{k-2}^{\mu \nu }C_{\mu \nu }^{*}+\bar{M}_{k-1}^{\mu \nu |\alpha }\eta _{\mu
\nu |\alpha }^{*}+\bar{N}_{k}^{\mu \nu |\alpha \beta }t_{\mu \nu |\alpha
\beta }^{*}\right. \right.  \nonumber \\
&&\left. \left. +\frac{1}{9}\tilde{S}_{k+1}^{\mu \nu \rho |\alpha \beta
\gamma }F_{\mu \nu \rho |\alpha \beta \gamma }\right) \right] +\partial
_{\mu }\psi _{k}^{\mu }.  \label{A107}
\end{eqnarray}
We observe that all the terms from the integrand are invariant. In order to
prove that the current $\psi _{k}^{\mu }$ can also be taken invariant, we
switch (\ref{A107}) to the original form notation
\begin{equation}
\alpha _{k}^{D}=\delta \lambda _{k+1}^{D}+d\lambda _{k}^{D-1},  \label{A107b}
\end{equation}
(where $\lambda _{k}^{D-1}$ is dual to $\psi _{k}^{\mu }$). As $\alpha
_{k}^{D}$ is by assumption invariant and we have shown that $\lambda
_{k+1}^{D}$ can be taken invariant, (\ref{A107b}) becomes
\begin{equation}
\beta _{k}^{D}=d\lambda _{k}^{D-1}.  \label{A107a}
\end{equation}
It states that the invariant polynomial $\beta _{k}^{D}=\alpha
_{k}^{D}-\delta \lambda _{k+1}^{D}$, of form degree $D$ and of strictly
positive antighost number, is $d$-exact. Then, in agreement with the Theorem
\ref{hginv} in form degree $D$ (see the paragraph following this theorem),
we can take $\lambda _{k}^{D-1}$ (or, which is the same, $\psi _{k}^{\mu }$)
to be invariant.

In conclusion, the induction hypothesis for antighost number ($k+3$) and
form degree $D$ leads to the same property for antighost number $k$ and form
degree $D$, which proves the theorem for all $k\geq 3$ since we have shown
that it holds for $k\geq D+3$. $\blacksquare $

The most important consequence of the last theorem is the validity of the
result (\ref{r81c}) on the triviality of $H^{\mathrm{inv}}\left( \delta
|d\right) $ in antighost number strictly greater than three.

\section{Local cohomology of $s$, $H\left( s|d\right) $\label{elim4}}

Now, we have all the necessary tools for the study of the local cohomology $%
H\left( s|d\right) $ in form degree $D$ ($D\geq 5$). We will show that it is
always possible to remove the components of antighost number strictly
greater than three from any co-cycle of $H_{D}^{g}\left( s|d\right) $ in
form degree $D$ only by trivial redefinitions.

We consider a co-cycle from $H_{D}^{g}\left( s|d\right) $, $sa+db=0$, with $%
\deg \left( a\right) =D$, $\mathrm{gh}\left( a\right) =g$, $\deg \left(
b\right) =D-1$, $\mathrm{gh}\left( b\right) =g+1$. Trivial redefinitions of $%
a$ and $b$ mean the simultaneous transformations $a\rightarrow a+sc+de$ and $%
b\rightarrow b+df+se$. We expand $a$ and $b$ according to the antighost
number and ask that $a_{0}$ is local, such that each expansion stops at some
finite antighost number \cite{gen2}, $a=\sum\nolimits_{k=0}^{I}a_{k}$, $%
b=\sum\nolimits_{k=0}^{M}b_{k}$, $\mathrm{agh}\left( a_{k}\right) =k=\mathrm{%
agh}\left( b_{k}\right) $. Due to (\ref{r45}), the equation $sa+db=0$ is
equivalent to the tower of equations
\begin{eqnarray*}
\delta a_{1}+\gamma a_{0}+db_{0} &=&0, \\
&&\vdots \\
\delta a_{I}+\gamma a_{I-1}+db_{I-1} &=&0, \\
&&\vdots
\end{eqnarray*}
The form of the last equation depends on the values of $I$ and $M$, but we
can assume, without loss of generality, that $M=I-1$. Indeed, if $M>I-1$,
the last $\left( M-I\right) $ equations read as $db_{k}=0$, $I<k\leq M$,
which imply that $b_{k}=df_{k}$, $\deg \left( f_{k}\right) =D-2$. We can
thus absorb all the pieces $\left( df_{k}\right) _{I<k\leq M}$ in a trivial
redefinition of $b$, such that the new ``current'' stops at antighost number
$I$. Accordingly, the bottom equation becomes $\gamma a_{I}+db_{I}=0$, so
the Corollary \ref{gaplusdb} ensures that we can make a redefinition $%
a_{I}\rightarrow a_{I}-d\rho _{I}$ such that $\gamma \left( a_{I}-d\rho
_{I}\right) =0$. Meanwhile, the same corollary (see the formula (\ref{A71a}%
)) leads to $b_{I}=dg_{I}+\gamma \rho _{I}$, where $\deg \left( \rho
_{I}\right) =D-1$, $\deg \left( g_{I}\right) =D-2$, $\mathrm{agh}\left( \rho
_{I}\right) =\mathrm{agh}\left( g_{I}\right) =I$, $\mathrm{gh}\left( \rho
_{I}\right) =g$, $\mathrm{gh}\left( g_{I}\right) =g+1$. Then, it follows
that we can make the trivial redefinitions $a\rightarrow a-d\rho _{I}$ and $%
b\rightarrow b-dg_{I}-s\rho _{I}$, such that the new ``current'' stops at
antighost number $\left( I-1\right) $, while the last component of the
co-cycle from $H_{D}^{g}\left( s|d\right) $ is $\gamma $-closed.

In consequence, we obtained the equation $sa+db=0$, with
\begin{equation}
a=\sum\limits_{k=0}^{I}a_{k},\;b=\sum\limits_{k=0}^{I-1}b_{k},  \label{exp}
\end{equation}
where $\mathrm{agh}\left( a_{k}\right) =k$ for $0<k<I$ and $\mathrm{agh}%
\left( b_{k}\right) =k$ for $0<k<I-1$. All $a_{k}$ are $D$-forms of ghost
number $g$ and all $b_{k}$ are $\left( D-1\right) $-forms of ghost number $%
\left( g+1\right) $, with $\mathrm{pgh}\left( a_{k}\right) =g+k$ for $0<k<I$
and $\mathrm{pgh}\left( b_{k}\right) =g+k+1$ for $0<k<I-1$. The equation $%
sa+db=0$ is now equivalent with the tower of equations (where some $\left(
b_{k}\right) _{0\leq k\leq I-1}$ could vanish)
\begin{eqnarray}
\delta a_{1}+\gamma a_{0}+db_{0} &=&0,  \label{A110} \\
&&\vdots  \nonumber \\
\delta a_{k+1}+\gamma a_{k}+db_{k} &=&0,  \label{A110ab} \\
&&\vdots  \nonumber \\
\delta a_{I}+\gamma a_{I-1}+db_{I-1} &=&0,  \label{A109} \\
\gamma a_{I} &=&0.  \label{A108}
\end{eqnarray}
\emph{Next, we show that we can eliminate all the terms }$\left(
a_{k}\right) _{k>3}$\emph{\ and }$\left( b_{k}\right) _{k>2}$\emph{\ from
the expansions (\ref{exp}) by trivial redefinitions only.}

Assuming that $a$ stops at an odd value of the pure ghost number, $g+I=2L+1$%
, the bottom equation, (\ref{A108}), yields $a_{I}\in H^{2L+1}\left( \gamma
\right) $. Then, in agreement with the result (\ref{hgzero}), $a_{I}$ is $%
\gamma $-trivial, $a_{I}=\gamma \bar{a}_{I}$, where $\mathrm{agh}\left( \bar{%
a}_{I}\right) =I$, $\mathrm{pgh}\left( \bar{a}_{I}\right) =g+2L$ and $\deg
\left( \bar{a}_{I}\right) =D$. Consequently, we can make the trivial
redefinition $a\rightarrow a-s\bar{a}_{I}$, whose decomposition stops at
antighost number $\left( I-1\right) $, such that the bottom equation
corresponding to the redefined co-cycle of $H_{D}^{g}\left( s|d\right) $
takes the form $\gamma a_{I-1}+db_{I-1}=0$. Now, we apply again the
Corollary \ref{gaplusdb} and replace it with the equation $\gamma a_{I-1}=0$%
, such that the new ``current'' can be made to end at antighost number $%
\left( I-2\right) $, $b=\sum\limits_{k=0}^{I-2}b_{k}$. In conclusion, if $%
g+I=2L+1$ is odd, we can always remove the last components $a_{I}$ and $%
b_{I-1}$ from a co-cycle $a\in H_{D}^{g}\left( s|d\right) $ and its
corresponding ``current'' by trivial redefinitions only.

We can thus assume, without loss of generality, that any co-cycle $a$ from $%
H_{D}^{g}\left( s|d\right) $ can be taken to stop at a value $I$ of the
antighost number such that $g+I=2L$, $a=\sum\limits_{k=0}^{I}a_{k}$, $%
b=\sum\limits_{k=0}^{I-1}b_{k}$. We consider that $I>3$. The last equation
from the system equivalent with $sa+db=0$ takes the form (\ref{A108}), with $%
\mathrm{pgh}\left( a_{I}\right) =g+I=2L$, so $a_{I}\in H^{2L}\left( \gamma
\right) $. In agreement with the general results on $H\left( \gamma \right) $
(see Subsection \ref{hgama}) it follows that
\begin{equation}
a_{I}=\stackrel{(0)}{a}_{I}+\cdots +\stackrel{(L)}{a}_{I}+\gamma \bar{a}_{I},
\label{A110a}
\end{equation}
where
\begin{equation}
\stackrel{(i)}{a}_{I}=\sum_{J}\alpha _{J,i}e^{J,i},\;i=0,\cdots ,L.
\label{A110b}
\end{equation}
All $\alpha _{J,i}$ are invariant polynomials, with
\begin{equation}
\mathrm{agh}\left( \alpha _{J,i}\right) =I,\;\deg \left( \alpha
_{J,i}\right) =D,  \label{A110c}
\end{equation}
and $e^{J,i}$ are the elements of pure ghost number $2L$ of a basis of
polynomials in $C_{\mu \nu }$ and $\partial _{\left[ \alpha \right.
}C_{\left. \mu \right] \nu }$ with the $\bar{D}$-degree equal to $i$ (see
the formula (\ref{A46f}) with $l$ replaced by $L$). Applying $\gamma $ on (%
\ref{A109}) and using (\ref{A108}), $\gamma ^{2}=0$ and $\gamma d+d\gamma =0$%
, we find that $-d\left( \gamma b_{I-1}\right) =0$, such that the triviality
of the cohomology of $d$ implies that
\begin{equation}
\gamma b_{I-1}+dc_{I-1}=0,  \label{A111}
\end{equation}
where $\mathrm{agh}\left( c_{I-1}\right) =I-1$, $\mathrm{pgh}\left(
c_{I-1}\right) =2L+1$, $\deg \left( c_{I-1}\right) =D-2$. From the Corollary
\ref{gaplusdb} it follows (as $I>3$ by assumption, so $I-1>0$) that we can
make a trivial redefinition such that (\ref{A111}) is replaced with the
equation
\begin{equation}
\gamma b_{I-1}=0.  \label{A112}
\end{equation}
In agreement with (\ref{A112}), $b_{I-1}$ belongs to $H^{2L}\left( \gamma
\right) $, so we can take
\begin{equation}
b_{I-1}=\stackrel{(0)}{b}_{I-1}+\cdots +\stackrel{(L)}{b}_{I-1}+\gamma \bar{b%
}_{I-1},  \label{A113}
\end{equation}
where
\begin{equation}
\stackrel{(i)}{b}_{I-1}=\sum_{J}\beta _{J,i}e^{J,i},\;i=0,\cdots ,L.
\label{A114}
\end{equation}
All $\beta _{J,i}$ are invariant polynomials, with
\begin{equation}
\mathrm{agh}\left( \beta _{J,i}\right) =I-1,\;\deg \left( \beta
_{J,i}\right) =D-1,  \label{A115}
\end{equation}
and $e^{J,i}$ are the elements of pure ghost number $2L$ of a basis of
polynomials in $C_{\mu \nu }$ and $\partial _{\left[ \alpha \right.
}C_{\left. \mu \right] \nu }$ with the $\bar{D}$-degree equal to $i$.
Inserting (\ref{A110a}--\ref{A110b}) and (\ref{A113}--\ref{A114}) in (\ref
{A109}) and using the fact that all the elements $e^{J,i}$ are commuting,
together with the relation (\ref{eqnew1}) for $b_{I-1}\in H^{2L}\left(
\gamma \right) $, we get that
\begin{equation}
\sum_{i=0}^{L}\sum_{J}\left[ \left( \delta \alpha _{J,i}+\bar{D}\beta
_{J,i}\right) e^{J,i}+\beta _{J,i}\bar{D}e^{J,i}\right] =\gamma \left(
-a_{I-1}-\hat{b}_{I-1}+\delta \bar{a}_{I}+d\bar{b}_{I-1}\right) ,
\label{A116}
\end{equation}
where $\hat{b}_{I-1}$ comes from $db_{I-1}=\bar{D}b_{I-1}+\gamma \hat{b}%
_{I-1}$. As $\delta \alpha _{J,i}$ and $\bar{D}\beta _{J,i}=d\beta _{J,i}$
are invariant polynomials, while $\bar{D}e^{J,i}=\sum_{J^{\prime
}}A_{J^{\prime },i+1}^{J,i}e^{J^{\prime },i+1}$ (see the formula (\ref{A40H}%
)), the property (\ref{A36}) ensures that the left-hand side of (\ref{A116})
must vanish
\begin{equation}
\sum_{i=0}^{L}\sum_{J}\left[ \left( \delta \alpha _{J,i}+\bar{D}\beta
_{J,i}\right) e^{J,i}+\beta _{J,i}\bar{D}e^{J,i}\right] =0.  \label{A117}
\end{equation}
Using the decomposition (\ref{A42}) and the definitions (\ref{A43a}--\ref
{A43h}), the projection of the equation (\ref{A117}) on the various values
of the $\bar{D}$-degree becomes equivalent with the equations
\begin{eqnarray}
0 &:&\delta \alpha _{J,0}+d\beta _{J,0}=0,  \label{A118a} \\
1 &:&\delta \alpha _{J,1}+d\beta _{J,1}+\beta _{J^{\prime
},0}A_{J,1}^{J^{\prime },0}=0,  \label{A118b} \\
&&\vdots  \label{A118c} \\
L &:&\delta \alpha _{J,L}+d\beta _{J,L}+\beta _{J^{\prime
},L-1}A_{J,L}^{J^{\prime },L-1}=0,  \nonumber
\end{eqnarray}
while the equation (\ref{A117}) projected on the value $\left( L+1\right) $
of the $\bar{D}$-degree is automatically satisfied, $\bar{D}_{1}e^{J,L}=0$,
since $\bar{D}_{1}\partial _{\left[ \alpha \right. }C_{\left. \mu \right]
\nu }=0$ and $e^{J,L}$ contains $L$ factors of the type $\partial _{\left[
\alpha \right. }C_{\left. \mu \right] \nu }$.

From (\ref{A118a}) we read that for all $J$ the invariant polynomials $%
\alpha _{J,0}$ belong to $H_{I}^{D}\left( \delta |d\right) $. Thus, as we
assumed that $I>3$ and we know that $H_{I}^{D}\left( \delta |d\right) =0$
for $I>3$, we deduce that all $\alpha _{J,0}$ are trivial
\begin{equation}
\alpha _{J,0}=\delta \lambda _{I+1,J,0}^{D}+d\lambda _{I,J,0}^{D-1},
\label{A119}
\end{equation}
where all $\lambda _{I+1,J,0}^{D}$ are $D$-forms of antighost number $\left(
I+1\right) $ and all $\lambda _{I,J,0}^{D-1}$ are $\left( D-1\right) $ forms
of antighost number $I$. Applying the result of the Theorem \ref{hinvdelta},
we have that all $\lambda _{I+1,J,0}^{D}$ and $\lambda _{I+1,J,0}^{D}$ can
be taken to be invariant polynomials, so all $\alpha _{J,0}$ are in fact
trivial in $H_{I}^{\mathrm{inv}D}\left( \delta |d\right) $. Replacing (\ref
{A119}) in (\ref{A118a}) and using $\delta ^{2}=0$ together with $\delta
d+d\delta =0$, we obtain that $d\left( -\delta \lambda _{I,J,0}^{D-1}+\beta
_{J,0}\right) =0$. As $\lambda _{I,J,0}^{D-1}$ and $\beta _{J,0}$ are
invariant polynomials of strictly positive antighost number and of form
degree $\left( D-1\right) $, by Theorem \ref{hginv} it follows that $-\delta
\lambda _{I,J,0}^{D-1}+\beta _{J,0}=d\lambda _{I-1,J,0}^{D-2}$, where $%
\lambda _{I-1,J,0}^{D-2}$ are also invariant polynomials for all $J$, with $%
\mathrm{agh}\left( \lambda _{I-1,J,0}^{D-2}\right) =I-1$ and $\deg \left(
\lambda _{I-1,J,0}^{D-2}\right) =D-2$, so
\begin{equation}
\beta _{J,0}=\delta \lambda _{I,J,0}^{D-1}+d\lambda _{I-1,J,0}^{D-2}.
\label{A120}
\end{equation}
From (\ref{A119}), we have that
\begin{eqnarray}
\stackrel{(0)}{a}_{I} &=&\sum_{J}\left( \delta \lambda
_{I+1,J,0}^{D}+d\lambda _{I,J,0}^{D-1}\right) e^{J,0}  \nonumber \\
&=&s\left( \sum_{J}\lambda _{I+1,J,0}^{D}e^{J,0}\right) +d\left(
\sum_{J}\lambda _{I,J,0}^{D-1}e^{J,0}\right) -\sum_{J}\left( \lambda
_{I,J,0}^{D-1}de^{J,0}\right) .  \label{A121}
\end{eqnarray}
As $de^{J,0}=\sum_{J^{\prime }}A_{J^{\prime },1}^{J,0}e^{J^{\prime
},1}+\gamma \hat{e}^{J,0}$ and $\gamma \lambda _{I,J,0}^{D-1}=0$, we find
that
\begin{eqnarray}
\stackrel{(0)}{a}_{I} &=&s\left( \sum_{J}\lambda
_{I+1,J,0}^{D}e^{J,0}\right) +d\left( \sum_{J}\lambda
_{I,J,0}^{D-1}e^{J,0}\right)  \nonumber \\
&&-\gamma \left( \sum_{J}\lambda _{I,J,0}^{D-1}\hat{e}^{J,0}\right)
-\sum_{J,J^{\prime }}\left( \lambda _{I,J,0}^{D-1}A_{J^{\prime
},1}^{J,0}e^{J^{\prime },1}\right) .  \label{A122}
\end{eqnarray}
Similarly, relying on (\ref{A120}) we deduce that
\begin{eqnarray}
\stackrel{(0)}{b}_{I-1} &=&s\left( \sum_{J}\lambda
_{I,J,0}^{D-1}e^{J,0}\right) +d\left( \sum_{J}\lambda
_{I-1,J,0}^{D-2}e^{J,0}\right)  \nonumber \\
&&-\gamma \left( \sum_{J}\lambda _{I-1,J,0}^{D-2}\hat{e}^{J,0}\right)
-\sum_{J,J^{\prime }}\left( \lambda _{I-1,J,0}^{D-2}A_{J^{\prime
},1}^{J,0}e^{J^{\prime },1}\right) .  \label{A123}
\end{eqnarray}
If we perform the trivial redefinitions
\begin{eqnarray}
a_{I}^{\prime } &=&a_{I}-s\left( \sum_{J}\lambda
_{I+1,J,0}^{D}e^{J,0}\right) -d\left( \sum_{J}\lambda
_{I,J,0}^{D-1}e^{J,0}\right) ,  \label{A124a} \\
b_{I-1}^{\prime } &=&b_{I-1}-s\left( \sum_{J}\lambda
_{I,J,0}^{D-1}e^{J,0}\right) -d\left( \sum_{J}\lambda
_{I-1,J,0}^{D-2}e^{J,0}\right) ,  \label{A124b}
\end{eqnarray}
and meanwhile partially fix $\bar{a}_{I}$ and $\bar{b}_{I-1}$ from (\ref
{A110a}) and respectively (\ref{A113}) to
\begin{eqnarray}
\bar{a}_{I} &=&\sum_{J}\lambda _{I,J,0}^{D-1}\hat{e}^{J,0}+\cdots ,
\label{A124c} \\
\bar{b}_{I-1} &=&\sum_{J}\lambda _{I-1,J,0}^{D-2}\hat{e}^{J,0}+\cdots ,
\label{A124d}
\end{eqnarray}
then (\ref{A122}--\ref{A123}) ensure that the lowest value of the $\bar{D}$%
-degree in the decompositions of $a_{I}^{\prime }$ and $b_{I-1}^{\prime }$
is equal to one. In conclusion, under the hypothesis that $I>3$, we
annihilated all the pieces from $a_{I}$ and $b_{I-1}$ with the $\bar{D}$%
-degree equal to zero by trivial redefinitions only. We can then
successively remove the terms of higher $\bar{D}$-degree from $a_{I}$ and $%
b_{I-1}$ by a similar procedure (and also the residual $\gamma $-exact terms
by conveniently fixing the pieces ``$\cdots $'' from $\bar{a}_{I}$ and $\bar{%
b}_{I-1}$) until we completely discard $a_{I}$ and $b_{I-1}$. Next, we pass
to a co-cycle $a$ from $H_{D}^{g}\left( s|d\right) $ that ends at the value $%
\left( I-1\right) $ of the antighost number, and hence $g+I-1=2L-1$ is odd,
so we can apply the arguments preceding the equation (\ref{A110a}) and
remove both $a_{I-1}$ and $b_{I-2}$. This two-step procedure can be
continued until we reach antighost number three. If $g+3$ is even we cannot
go down and discard $a_{3}$ and $b_{2}$, since both $H^{g+3}\left( \gamma
\right) $ and $H_{3}^{D\mathrm{inv}}\left( \delta |d\right) $ are
non-trivial. However, if $g+3$ is odd, then $H^{g+3}\left( \gamma \right) =0$%
, so we can go one step lower and remove $a_{3}$ and $b_{2}$. In conclusion,
we can take, without loss of generality
\begin{eqnarray}
a &=&a_{0}+a_{1}+a_{2}+a_{3},\;b=b_{0}+b_{1}+b_{2},\;\mathrm{if}\;g+3=2l,
\label{A125} \\
a &=&a_{0}+a_{1}+a_{2},\;b=b_{0}+b_{1},\;\mathrm{if}\;g+3=2l+1,  \label{A126}
\end{eqnarray}
in the equation $sa+db=0$, where $\mathrm{gh}\left( a\right) =g$.
Furthermore, the last terms can be assumed to involve only non-trivial
elements from $H^{\mathrm{inv}}\left( \delta |d\right) $.

\section{Conclusion}

To conclude with, in this paper we have used some specific cohomological
techniques, based on the Lagrangian BRST differential, to prove that any
non-trivial co-cycle from the local BRST cohomology in form degree $D$ for a
free, massless tensor field $t_{\lambda \mu \nu |\alpha }$, can be taken to
stop at antighost number three, its last component belonging to $H\left(
\gamma \right) $ and containing a non-trivial element from $H^{\mathrm{inv}%
}\left( \delta |d\right) $. This result is based on various cohomological
properties involving the exterior longitudinal derivative, the Koszul-Tate
differential, as well as the exterior spacetime differential, which have
been proved in detail. The issues addressed in this paper are important from
the perspective of constructing consistent interactions for this type of
mixed symmetry tensor field since it is known that the first-order
deformation of the solution to the master equation is a co-cycle of the
local BRST cohomology $H_{D}^{0}\left( d|s\right) $ in form degree $D$ and
ghost number zero.

\section*{Acknowledgment}

The authors thank Professor Gennadi Sardanashvily for inviting them to
contribute with this paper to the special issue of International Journal of
Geometric Methods in Modern Physics dedicated to the subject of ``Geometry
of Gauge Fields (50 years of gauge theory)''. The authors also thank Nicolas
Boulanger for valuable suggestions and comments. This work has been
supported from a type A grant with the Romanian National Council for
Academic Scientific Research and the Romanian Ministry of Education,
Research and Youth.

\end{document}